\documentclass[acmtog]{acmart}

\AtBeginDocument{%
  \providecommand\BibTeX{{%
    \normalfont B\kern-0.5em{\scshape i\kern-0.25em b}\kern-0.8em\TeX}}}

\acmJournal{TOG}
\acmArticle{261}
\acmVolume{44}
\acmNumber{6}
\acmMonth{12}
\acmPrice{}\acmDOI{10.1145/3763309}
\usepackage{amsfonts}
\usepackage{amsmath}
\usepackage{multirow}
\usepackage{booktabs}
\usepackage{enumerate}
\usepackage{graphicx} %
\usepackage{caption} %
\usepackage{multicol} %
\usepackage{xspace}
\usepackage{cleveref}
\usepackage{pifont}
\usepackage{wrapfig}

\newcommand{\revision}[1]{\textcolor{blue}{#1}}

\begin{document}

\title{Template-free and Rig-free Animation Transfer using Kinetic Codes}

\author{Sanjeev Muralikrishnan}
\authornote{Both authors contributed equally to this research.}
\email{sanjeev.muralikrishnan.20@ucl.ac.uk}
\affiliation{%
  \institution{University College London}
  \country{UK}
  \institution{Dolby Laboratories}
  \country{India}
}

\author{Niladri Shekhar Dutt}
\authornotemark[1]
\email{niladri.dutt.22@ucl.ac.uk}
\affiliation{%
  \institution{University College London}
  \country{UK}
}

\author{Niloy J. Mitra}
\email{n.mitra@cs.ucl.ac.uk}
\affiliation{%
  \institution{University College London}
  \country{UK}
  \institution{Adobe Research}
  \country{UK}
}

\newcommand{\name}{SMF\xspace}
\newcommand{\shape}{$X$\xspace}
\newcommand{\initshape}{\ensuremath{X_0}\xspace}
\newcommand{\shapet}{$X_t$\xspace}
\newcommand{\step}{$t$\xspace}
\newcommand{\sig}{\ensuremath{m_z}\xspace}
\newcommand{\motionT}{\ensuremath{M_t}\xspace}
\newcommand{\JposeT}{\ensuremath{J^P_t}\xspace}
\newcommand{\Jposefirst}{\ensuremath{J^P_0}\xspace}
\newcommand{\features}{\ensuremath{C}\xspace}
\newcommand{\Jresidualfirst}{\ensuremath{J^R_0}\xspace}
\newcommand{\JresidualT}{\ensuremath{J^R_t}\xspace}
\newcommand{\Jt}{\ensuremath{J_t}\xspace}
\newcommand{\funcPose}{\ensuremath{f_P\xspace}}
\newcommand{\funcResidual}{\ensuremath{f_R\xspace}}
\newcommand{\attnPoseW}{\ensuremath{E^{P}_{W_n}\xspace}}
\newcommand{\attnResidualPast}{\ensuremath{E^C_{W_{n-1}}\xspace}}
\newcommand{\JposeW}{\ensuremath{J_{k:k+W}}\xspace}
\newcommand{\JresidualWpast}{\ensuremath{J^C_{k:k-W}}\xspace}
\newcommand{\timeW}{\ensuremath{T_{k:k+W}}\xspace}
\newcommand{\timeWpast}{\ensuremath{T_{k:k-W}}\xspace}

\newcommand{\vnudge}{\vspace*{-.15in}}

\newcommand{\shapesig}{\ensuremath{\beta}\xspace}
\newcommand{\shapesigPred}{\ensuremath{\beta'}\xspace}
\newcommand{\betaloss}{\ensuremath{L_{\beta}}\xspace}
\newcommand{\common}{\ensuremath{G}\xspace}
\newcommand{\centroid}{\ensuremath{\mathbf{c}}\xspace}
\newcommand{\stateForA}{\ensuremath{e_{acc}}\xspace}
\newcommand{\stateForV}{\ensuremath{e_{vel}}\xspace}
\newcommand{\stateA}{\ensuremath{s_{i_{acc}}}\xspace}
\newcommand{\stateV}{\ensuremath{s_{i_{vel}}}\xspace}
\newcommand{\jointsT}{\ensuremath{R_t}\xspace}
\newcommand{\funcAcc}{\ensuremath{f_\text{acc}^J}\xspace}
\newcommand{\loss}{\ensuremath{\mathcal{L}}\xspace}

\newcommand{\vode}{VertexODE\xspace}
\newcommand{\njf}{NJF\xspace}
\newcommand{\softSmpl}{SoftSMPL\xspace}

\newcommand{\inpmotion}{\ensuremath{\gamma_k}\xspace}
\newcommand{\shortname}{\ensuremath{SMF}\xspace}
\newcommand{\deform}{\ensuremath{f_D}\xspace}
\newcommand{\correct}{\ensuremath{f_C}\xspace}
\newcommand{\firstparam}{\ensuremath{\gamma_0}\xspace}
\newcommand{\deltaparam}{\ensuremath{\gamma^\delta_k}\xspace}
\newcommand{\deltasequence}{\ensuremath{\{\gamma^\delta_1,\ldots,\gamma^\delta_k,\ldots,\gamma^\delta_{N_f}\}}\xspace}
\newcommand{\predparam}{\ensuremath{\hat{\gamma}_k}\xspace}
\newcommand{\encoder}{\ensuremath{\mathcal{E}_\gamma}\xspace}
\newcommand{\decoder}{\ensuremath{\mathcal{D}_\gamma}\xspace}
\newcommand{\latent}{\ensuremath{z_k}\xspace}
\newcommand{\deformjac}{\ensuremath{J_k}\xspace}
\newcommand{\initjac}{\ensuremath{J_0}\xspace}
\newcommand{\deformjacres}{\ensuremath{J^R_k}\xspace}
\newcommand{\correctivek}{\ensuremath{J^C_k}\xspace}
\newcommand{\correctivefirst}{\ensuremath{J^C_0}\xspace}
\newcommand{\correctivefirstaug}{\ensuremath{\hat{J}^C_0}\xspace}
\newcommand{\correctivekaug}{\ensuremath{\hat{J}^C_k}\xspace}
\newcommand{\project}{\ensuremath{W_p}\xspace}
\newcommand{\numjoints}{\ensuremath{N_{joints}}\xspace}
\newcommand{\numframes}{\ensuremath{N_{f}}\xspace}

\begin{abstract}

Animation retargetting applies sparse motion description (e.g., keypoint sequences) to a character mesh to produce a semantically plausible and temporally coherent full-body mesh sequence. %
Existing approaches come with restrictions -- they require 
access to template-based shape priors or artist-designed deformation rigs, suffer from limited generalization to unseen motion and/or shapes, or exhibit motion jitter. 
We propose Self-supervised Motion Fields (\name), a self-supervised framework that is trained with only sparse motion representations, without requiring dataset-specific annotations, templates, or rigs. 
At the heart of our method are Kinetic Codes, a novel autoencoder-based sparse motion encoding, that exposes a semantically rich latent space, simplifying large-scale training. 
Our architecture comprises dedicated  spatial and temporal gradient predictors, which are jointly trained in an end-to-end fashion. The combined network, regularized by the Kinetic Codes' latent space, has good generalization across both unseen shapes and new motions.
We evaluated our method on unseen motion sampled from AMASS, D4D, Mixamo, and raw monocular video for animation transfer on various characters with varying shapes and topology. We report a new SoTA on the AMASS dataset in the context of generalization to unseen motion. Code, weights, and supplementary are available on the project webpage at 
\href{https://motionfields.github.io/}{\revision{https://motionfields.github.io/}}.

\end{abstract}



\begin{teaserfigure}
  \includegraphics[width=\textwidth]{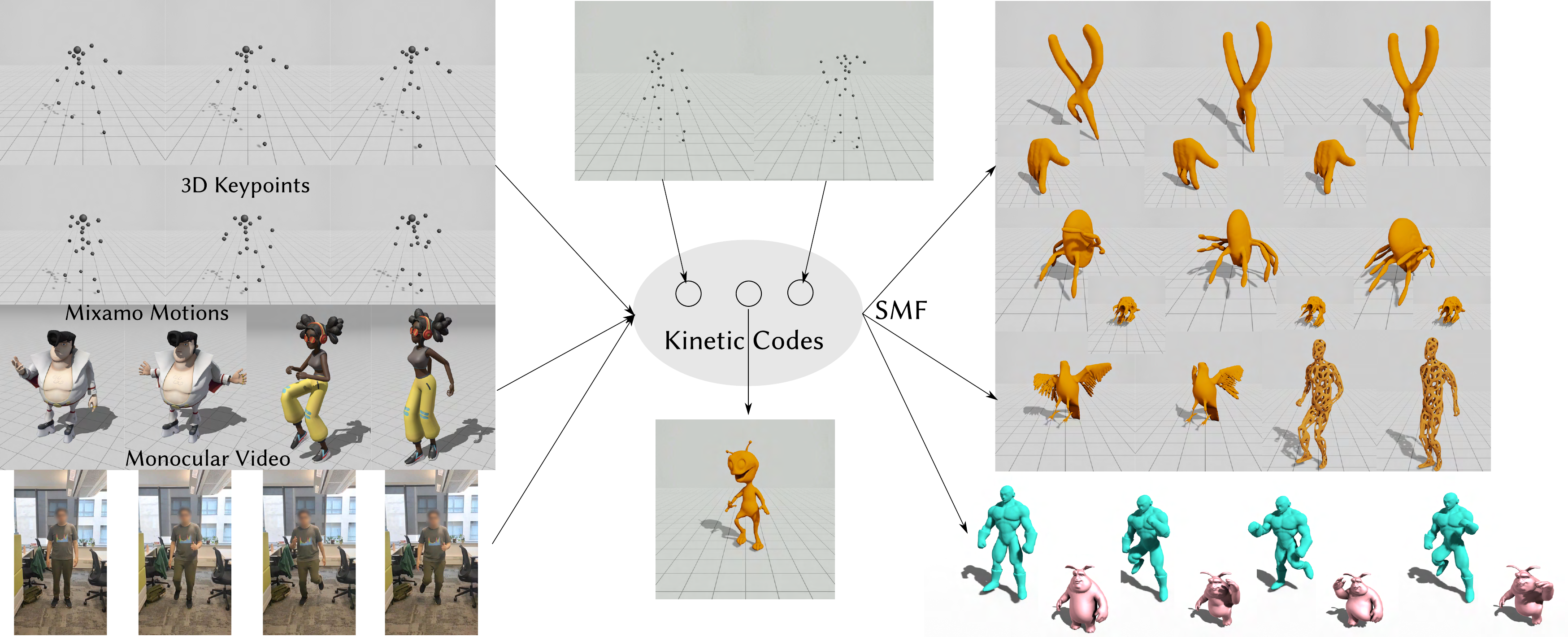}
  \caption{\textbf{Animation Transfer from Unseen Motion to Different Characters.} We present a self-supervised method to transfer coarse motion sequences, embedded in a learned Kinetic Code~(KC) space, to full body motion. Samples from the KC space can be consumed by our method, Self-supervised Motion Fields~(SMF), to produce mesh animations. Our method is trained with sparse signals and can be used for motion interpolation . We do not assume access to any morphable model,  canonical template mesh, or deformation rigs.   
Left shows various sparse motion inputs (3D keyframes, Mixamo sequences, or monocular video) that can be embedded into the learned KC space and decoded, and consumed by temporally coherent motion prediction via SMF to produce animations for different characters~(right), with varying topology and shapes. SMF can faithfully transfer human motion to non-humanoid characters. }
  \label{fig:teaser}
\end{teaserfigure}

\maketitle

\section{Introduction}

Motion brings characters to life. To build any digital twin world, be it for scenario planning, games, or movies, adding motion to static characters is a fundamental requirement.  Manually authoring full-body motion is tedious, error-prone, and requires significant effort from skilled artists. This quickly becomes expensive when scaling to long sequences or animating many different characters. Hence, researchers seek data-driven solutions.

Traditional approaches investigate this problem with explicit priors --- 
statistical body templates (e.g., SMPL~\cite{SMPL:2015} for humans, CAFM~\cite{cafm} for animals) to model shape and pose variations or constructing specific character rigs to transfer joint motion to full body motion via skinning weights. These approaches are simple, popular, and robust, but come at the cost of having to first build an expressive shape space along with a corresponding pose space, and are restricted to specific templates. %

Automating these workflows with learning-based solutions has gained popularity: learning a space of stick figure character motion (e.g., neural motion field~\cite{he2022nemf}) or phase-based character control~\cite{phaseFunctioned:17}); enabling deformation transfer from a source-target pair to a new shape via neural Jacobian fields~\cite{aigerman2022neural,trj}; or learning skinning weights from a set of annotated rigged characters~\cite{RigNet:20,qin2023NFR}. These approaches, however, require various levels of intermediate supervision, are limited in their handling of motion or shape variations, and suffer from limited generalization.

Given a coarse motion specification, we transfer animation to a full-character mesh, \textit{without} access to any rig or morphable template, at train or test times. A natural approach is to treat this as a sequence prediction problem. 
However, this quickly leads to memory issues as the animations' length or the meshes' resolution increases. Increasing the network capacity adversely affects the situation, leading to overfitting as we often have sparse/limited training data. Also, treating the problem at the frame level is efficient, but leads to jittery motion without any temporal coupling.

Inspired by the recent success of latent space diffusion models~\cite{rombach2021highresolution} over pixel space diffusion models, we ask if a similar latent space can be designed for (sparse) motion sequences. To this end, we propose \textit{Kinetic Codes}, a temporally-informed lightweight motion autoencoder, that we train over a collection of sparse (humanoid) motion sequences across all types of motion. (Since we only rely on keypoints, instead of body meshes/template, we call the latent space kinetic instead of kinematic.) 
Regularized by this latent space, we train a spatial and a temporal gradient predictor network. We couple the networks through differentiable spacetime integration and supervise the framework, in an end-to-end fashion.

By representing source motion using only keypoints, we eliminate the need for geometric constraints such as 2-manifoldness, watertightness, or fixed triangulation in source meshes. Moreover, our motion representation allows for 2D source motion as input, which can be transferred to any stylized character (see Figure~\ref{fig:teaser}). This keypoint-based representation simplifies motion capture and facilitates sampling and interpolation of motions. 

We evaluate \name for generalization across diverse shapes and unseen motion, and compare against various alternatives. We evaluate our setup on a range of diverse shapes and motion datasets (e.g., AMASS, D4D, Mixamo, monocular video). In summary, we: 
(i)~propose a self-supervised animation transfer framework regularized by kinetic codes, a learned temporally aware latent space; 
(ii)~develop a rig- and template-free animation transfer framework based on simple keypoints as input that is easy to train with sparse supervision and generalizes robustly to new motions and stylized characters; and
(iii)~report a new SoTA on the AMASS dataset and show realistic motion transfer to in-the-wild stylized characters using different 3D as well as 2D coarse space motion specifications. 

Code, weights, and supplementary are available on the project webpage at 
\href{https://motionfields.github.io/}{\revision{https://motionfields.github.io/}}.

\section{Related Works}

\paragraph*{Encoding shape deformation.} Parameterized deformation approaches represent 2D or 3D shapes through a predetermined function of shared parameters and capture deformations as variations of these parameters. A popular example of such models is morphable models~\cite{morphableSurvey:19}. Such techniques encompass cages, explicit formulations~\cite{ju2005meanvalue} or neural approaches~\cite{Yifan:NeuralCage:2020}, blendshapes~\cite{lewis2014blendshape}, skinned skeletons~\cite{jacobson2014skinning}, Laplacian eigenfunctions~\cite{rong2008spectral}, etc. Linking these parameters to the shapes' surface typically necessitates manual annotation of weights, commonly called weight painting, in 3D authoring tools. Alternatively, with access to sufficient supervision data, data-driven approaches can yield realistic neural rigs, as exemplified by Pinocchio~\cite{baran2007automatic}, RigNet~\cite{RigNet:20}, skinning-based motion retargeting~\cite{marsot2023correspondencefree, skeleton-aware2020, zhang2023skinned}, and skeletal articulations with neural blend shapes~\cite{li2021learning}. Unsupervised shape and pose disentanglement~\cite{zhou20unsupervised} proposes learning a disentangled latent representation of shape and pose, facilitating motion transfer using shape codes, dependent on registered meshes and maintaining identical connectivity. 
To plausibly animate these parameterized shapes over time, the parameters should evolve dynamically, weighing the mesh. Although these methodologies require access to body templates and/or rigs, they can still produce jittery results due to loose coupling of individual frame predictions. 

Notably, Skeleton-free pose transfer~\cite{skeletonfree} aims to alleviate the need for rigs by treating the character pose as a set of independent part deformations. By learning the skinning weights and deformations associated with each module, it can match the source pose using linear blending of skinning weights. While it achieves impressive results in pose transfer, being a per-frame method, it suffers from artifacts and lacks temporal coherence when applied to animation transfer, as observed in our results. Furthermore, its reliance on both source and target shapes being provided in a rest T-pose presents a practical limitation, as such canonical poses are often difficult to define for non-humanoids. This dependency challenges its classification as a truly rig-free method.

\paragraph*{Modeling motion as sequence prediction.}
Deep recurrent neural networks are capable of modeling time and shape sequences using  LSTMs to predict human joints~\cite{fragkiadaki2015erd}, generate motion in-betweening~\cite{harvey2020robust}, and to learn a motion field through time~\cite{he2022nemf}. %
These approaches require access to templates/rigs, and large datasets of joint motion since they are discrete time representations. Qiao et al.~\shortcite{qiao2018learning} utilize mesh convolutions with LSTMs to deform vertices through time; while,   Motion Diffusion~\cite{raab2023single} uses local attention to capture motifs of a single motion and combines it with a diffusion UNet module to produce motion extrapolation and in-betweening. 
The main challenge is handling long (extrapolation) sequences while still being able to generalize to unseen motion. 
In this work, we propose using Augmented Neural ODEs~\cite{dupont2019augmented}, operating through temporally-aware kinetic codes, to model time continuously instead of using discretized sequential networks such as LSTMs. 

\paragraph*{Modeling motion using morphable templates.}
Temporal surface effects can be modeled,  physically correctly, by simulating the underlying soft tissues with finite element methods~\cite{chadwick1989layered,fan2014active}. However, this direct simulation is often slow and requires artists to design the underlying bone and muscle structure~\cite{abdrashitov2021musculoskeletal}. To overcome the stiffness problem and speed up the simulation, reduced-order models have been proposed~\cite{modi2020efficient,park2008data}. 
When character rigs are available,  approaches have been proposed to add soft tissue deformation as an additive per-vertex bump map on top of a primary motion model. Santesteban et al.~\shortcite{santesteban} use this 
approach, AMASS~\cite{AMASS:ICCV:2019} imparts secondary motion using the blending coefficients of the SMPL shape space~\cite{SMPL:2015}, while Dyna~\cite{pons2015dyna} learns a data-driven model of soft-tissue deformations using a linear PCA subspace. However, these methods require access to primary motion via a skeleton rig and are restricted to humans registered to a canonical template.

Temporal Residual Jacobian~(TRJ)~\cite{trj} uses NJF and ODE for motion retargeting, and demonstrates good generalization across shapes. However, TRJ  requires motion annotation and also access to SMPL template during training, has to be retrained for different motion classes, produces jitters due to per-frame prediction, and does \textit{not} generalize to unseen motion. Unlike TRJ, which requires a template to learn motion transfer between different shapes, our self-supervised method reconstructs the original motion on the same shape, eliminating the need for annotated data.

\begin{figure*}[t!]
    \centering
    \includegraphics[width=\textwidth]{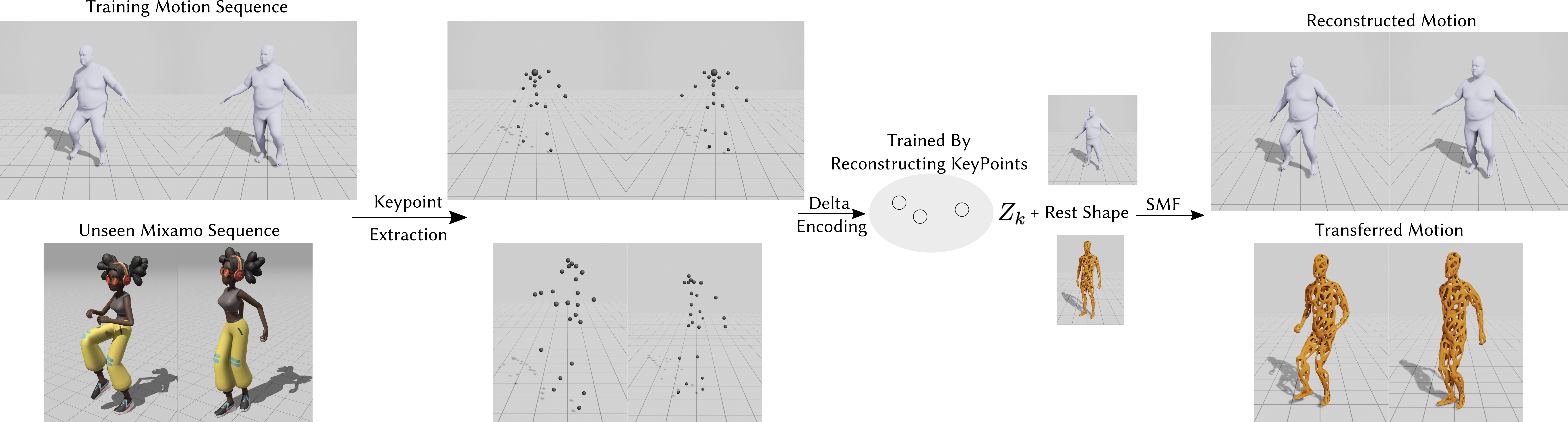} 
    \caption{{\textbf{Method overview.}} We present a self-supervised learning setup to transfer sparse motion information, specified in the form of keypoints over time, to target characters producing full-body motion. \textit{Top:} During training, given a motion dataset we extract sparse keypoints from the meshes and encode them to a novel \textit{Kinetic Code} representation. We then train two networks to map the rest shape and the Kinetic Code to the full body motion, with only mesh-level  reconstruction loss. \textit{Bottom:} At inference, we drop in stylized characters (\texttt{Hole Man}) and unseen motion inputs to obtain full-body character animation.}
    \label{fig:pipeline}
\end{figure*}

\section{Algorithm}
Our goal is to animate unrigged triangulated meshes of 3D characters, conditioned on coarse motion signal. These motion control parameters are defined at a few keypoints of the body, can be varied in representation (e.g., 2D or 3D), and specify the target pose per frame. We learn to map these coarse keypoint parameters to dense 3D meshes and, as output, generate an animated 3D mesh at every frame described by the input motion. This nullifies the need for any consistent  mesh template or fixed triangulation for both the source motion as well as the target character. 
Additionally, our method 
generalizes to unseen motion targets and  unseen body shapes, 
learns from sparse datasets containing a mix of motion examples,  and 
 can be applied to long motion sequences.

\subsection{Overview}
Our method is a module named Self-Supervised Motion Fields~(\shortname) that maps inputs describing a target shape, \initshape, and motion, $\{\inpmotion\}$, to per-frame motion as 3D meshes, 
\begin{equation}
\label{eq:blank_model}
    X_k := \shortname(\dots), 
\end{equation}
where $X_k$ denotes the mesh vertices at frame $k$.

\paragraph*{Input representation.} Our inputs are coarse motion parameters, defined as \inpmotion, which characterize the pose required in each frame $k$ and the target triangulated mesh defined as \initshape. In our experiments, we have tested $k$ ranging from 200 (short) to 4000 (long) sequences. We define \inpmotion at fixed keypoints extracted automatically from a given body. %
Specifically, \inpmotion is a $\numjoints \times D $ vector at each time step, where \numjoints is the number of keypoints/joints and $D$ is the dimensionality of the chosen input representation. This representation can be 3D keypoint locations on a mesh, 2D keypoints defined on a stick figure frame, or 3D relative Euler angles computed according to the kinematic tree of the chosen pose space. In this work, we focus on results with 2D and 3D keypoints, as we observe their performance to be more robust than 3D relative Euler angles. Our method is self-supervised as we automatically extract the keypoints from mesh sequence.  
\\
We now define our module from \Cref{eq:blank_model} as, 
\begin{equation}
    X_k := SMF(\inpmotion,\initshape,C),
\end{equation}
where $C$ describes additional geometric features of the target character such as centroids, normals, and Wave Kernel Signatures~\cite{aubry2011wave} computed on the faces of \initshape and encoded per-face using a shallow PointNet~\cite{qi2016pointnet}. These additional features establish correspondence during inference on unseen in-the-wild shapes. We jointly train the PointNet network and other networks in our system. %
Figure~\ref{fig:pipeline} presents an overview of our method. 

\paragraph{Keypoint extraction.}\label{subsec:extraction}
We semi-automatically extract the keypoints \inpmotion. We start with a one-time manual annotation of \textbf{one} human and \textbf{one} animal shape, performed by selecting mesh faces at joint locations. These sparse annotations are then automatically propagated across all characters and motions using point correspondences computed with Diff3F~\cite{dutt2023diffusion}. Our framework is 
 \setlength{\intextsep}{3pt}
 \begin{wrapfigure}[8]{r}{0.5\columnwidth}
     \centering
     \hskip -23pt\includegraphics[width=0.55\columnwidth]{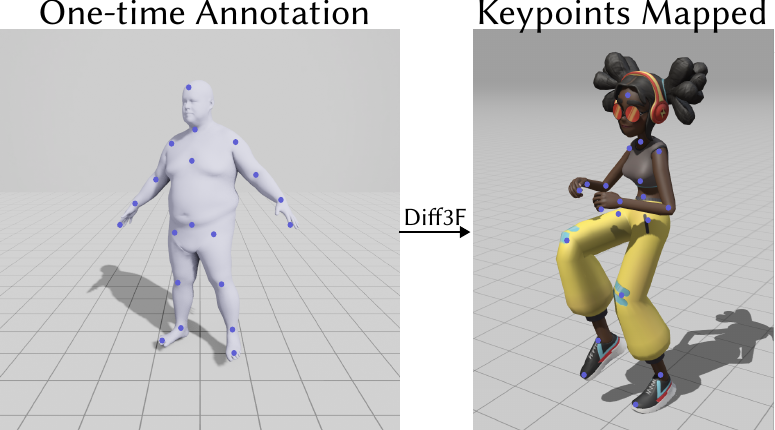}
     \Description{keypoint extraction}
     \label{fig:keypoint}
 \end{wrapfigure}
robust to variations between character setups; for example, it bridges the gap even when target keypoint locations (e.g., from Mixamo) differ from the source. This flexibility is possible because our method only requires the number and order of keypoints to be consistent, not their exact spatial positions. This process, where we annotate just two shapes once, allows us to robustly handle diverse character and motion datasets. For extracting 2D keypoints from RGB videos, we use HRNet~\cite{hrnet}, a pretrained pose extractor.

\subsection{\textit{Kinetic Codes}: Temporally-Informed Motion Representation}
\label{sec:temporalencoding}
At the core of our method is the shape deformation module \deform. We observed that naively passing the extracted coarse motion parameters \inpmotion to \deform impairs training, and leads to poor generalization and artifacts on unseen motions (see ablation in \Cref{sec:results}). This is unsurprising as an individual \inpmotion does \textit{not} contain any temporal motion information or context. Therefore, we first embed \inpmotion in the latent space of an autoencoder. By coupling information across time, this representation leads to smoother interpolation, thereby enhancing generalization to unseen motions. %

Further, given the sparsity of training data, we find that using the displacements of the motion parameters as inputs, instead of their absolute values, significantly boosts generalization to unseen motion. Therefore, we express  \inpmotion as displacement vectors with respect to the first frame motion parameter \firstparam as,
\begin{equation}
    \deltaparam := \inpmotion - \firstparam.
\end{equation}
We then train a multi-headed attention auto-encoder with self-attention to reconstruct \deltaparam as, 
\begin{eqnarray}
        \latent &=& \encoder(\deltaparam,\deltasequence) \label{eq:latent}\\
        \predparam &=& \decoder(\latent),
\end{eqnarray}
where \encoder, \decoder are multi-headed attention encoder and decoder networks, respectively; \latent is the per-frame latent motion representation, referred to as  \textit{kinetic code}, of the same dimensionality as \inpmotion, and \predparam are the decoded motion displacements; $N_f$ refers to number of frames.
\setlength{\intextsep}{1pt} %
\begin{wrapfigure}{r}{0.5\columnwidth} %
    \hspace*{0pt}\includegraphics[width=\linewidth]{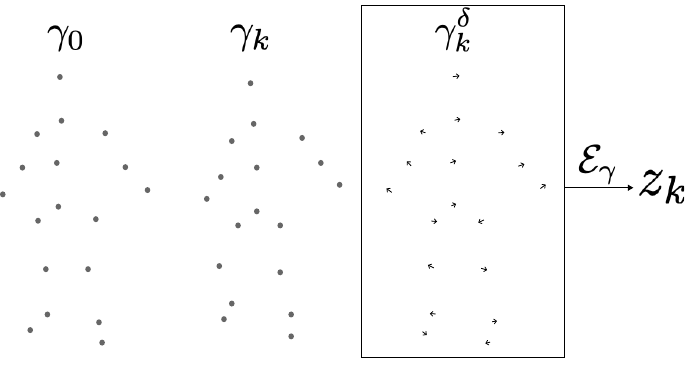}
    \Description{deltaencoding}
    \label{fig:deltaencoding}
\end{wrapfigure}
The length of the sequence can vary across motion samples. Note that we use attention on the full sequence as context, and hence a single frame \latent has an understanding of the broader motion sequence.

We train the auto-encoder by minimizing the reconstruction loss, 
\begin{equation}
    L_\gamma := \sum_{k=1}^{N_f} \|\predparam - \deltaparam\|^2.
\end{equation}
Thus, we obtain a temporally-informed latent motion representation \latent, which is more conducive to interpolation, resulting in improved generalization to unseen motions and unseen shapes. For our deformation module (described next), we only use \latent as the per-frame motion representation, and freeze \encoder and \decoder.

\subsection{3D Shape Posing via Deformation Module}
\label{sec:posing}
Inputs to our shape deformation module \deform comprise the initial configuration, specifically the shape \initshape in its canonical rest pose, the learned motion latents \latent, and the geometric features $C$ (derived from \initshape). The module \deform is designed to forecast the pose for each time step $k \in [1, \numframes]$. 
While \deform could be trained to directly forecast the target vertex positions, such an approach results in artifacts, including flipped and folded faces.
Following neural gradient space processing \cite{aigerman2022neural,trj}, we 
use a simple MLP to predict affine transformation matrices, labeled as Jacobians, at the centroids of the mesh faces, thus encoding relative transforms~\cite{defTransfer:sigg:04}. 
Specifically, we predict final vertex positions using predicted Jacobians, integrated through a differentiable Poisson solver to solve a system of linear equations. We supervise this using vertex-to-vertex and Jacobian losses to ensure accuracy in matching ground-truth data; this frees us from requiring additional annotation data. This improves shape consistency and can model various deformations by predicting affine transformations to mesh faces. %
Concretely, we define, 
\begin{eqnarray}
    \deformjacres &=& \deform(\initjac,\latent,C) \nonumber \\
    \deformjac &=& \initjac + \deformjacres = \initjac + \deform(\initjac,\latent,C),
    \label{eq:deformjac}
\end{eqnarray}
where \initjac is the Jacobian of the initial frame shape \initshape and \deformjacres are intermediate residuals predicted by \deform, which are added to \initjac to produce the Jacobian \deformjac for the $k^{th}$ frame. \setlength{\intextsep}{2pt} %
\begin{wrapfigure}{9}{0.5\columnwidth} %
    \hskip -10pt \includegraphics[width=\linewidth]{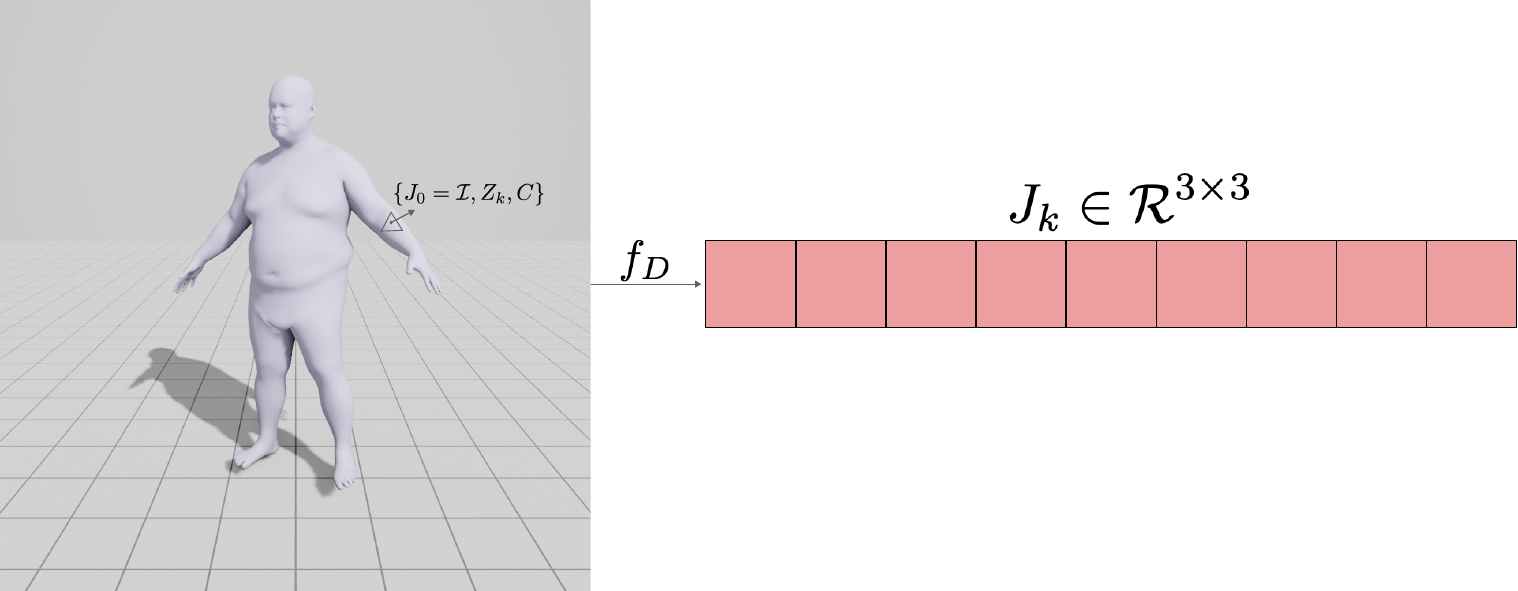}
    \Description{Per Frame Jacobians}
    \label{fig:perframejacobians}
\end{wrapfigure} We found that predicting the Jacobians using the Kinetic codes, coupled with residual connections leads to significantly faster and well-behaved convergence and better generalization. Note that \initjac is the identity transformation projected on the local basis of each face of \initshape by rotation. All Jacobians are predicted in the central coordinate frame defined by the local bases of the faces of \initshape.

We can now directly obtain final vertex positions using the Poisson Solver on the predicted Jacobians. However, as we drift away from the pose of the initial shape \initshape or when inferring on unseen motion parameters, accumulated error leads to shape inconsistencies and artifacts. Additionally, \deform utilizes only coarse motion parameters and, by itself, is unable to produce temporally coherent and smooth motions on dense meshes for long sequences. Hence, we use an Augmented ODE based formulation, described next, to enable temporally coherent predictions.

\subsection{Temporally Coherent Motion Prediction}
\label{sec:augode}

\paragraph*{Motion prediction.} 
We introduce a second stage that learns to predict a dynamic correction to the initial pose estimate to enforce temporal coherence. We predict the sequences in chunks of consecutive frames, i.e., the given sequence is split into fixed windows, each of size $W$ ($W=32$ in our tests). We set the initial state of the NODE as the Corrective Jacobians required for the first frame as, 
\begin{equation}
\label{eq:correctivek}
    \correctivefirst = \textbf{0} \in \mathbb{R}^{3 \times 3}.
\end{equation}
To make this operable with ANODE~\cite{dupont2019augmented}, we augment the initial state with extra dimensions. We follow the recommendations of the original work and set the extra dimensions to $\textbf{0} \in \mathbb{R}^A$ where $A$ denotes the number of extra (augmented) dimensions to be added. We use $A=256$ for all our experiments. In essence, these augmented dimensions act as a form of temporal memory for the ODE. By lifting the state space to a higher dimensional space where it can more easily predict the required trajectories, the system is able to maintain a richer state representation of the motion's history. This simplifies the task of learning the corrective dynamics needed to prevent drift. In other words, temporal message passing helps predict the corrective Jacobians. \Cref{eq:correctivek} becomes,
\begin{align}
\label{eq:correctivekaug}
    \correctivefirstaug = \textbf{0} \in \mathbb{R}^{(3 \times 3) + A}.
\end{align}
We predict the correctives in the augmented space first, which is driven by an MLP \correct that predicts the rate of change of the correctives in time. The function \correct models the rate of change of the drift based on conditioning factors,
\begin{equation}
    \frac{\partial \correctivekaug}{\partial t} = \\ \correct(\hat{J}_0,\attnPoseW,\attnResidualPast,t)
    \label{eq:funcResidual}
\end{equation}
where $J_0$, as defined previously, are the Jacobians of the first frame; \attnPoseW is the attention encoding of the current window of Jacobian predictions \deformjac from \Cref{eq:deformjac} and \attnResidualPast is the attention encoding of the previous window of corrective predictions. Intuitively, \attnPoseW provides the context of the current motion's structure, while \attnResidualPast informs the model about the accumulated error from the previous window. We integrate the local changes over time using Euler's method to obtain \correctivek at each time as,
\begin{equation}
    \correctivekaug = \int_0^t \frac{\partial \correctivekaug}{\partial t} dt + \correctivefirstaug 
=  \int_0^t \correct(\hat{J}_0,\attnPoseW,\attnResidualPast,t) dt + \correctivefirstaug.
    \label{eq:ode}
\end{equation}
Since \correctivekaug is in the augmented space, we use a final linear projection with learnable weights ($W_p$) to map it back to the original unaugmented space. Specifically, 
\begin{equation}
    \correctivek = \project\correctivekaug
    \label{eq:finalcorrective}
\end{equation}
where \project is simply the learnable weights of the last linear layer projecting from $(3 \times 3)+A$ to the $(3 \times 3)$ Jacobian.

Our jointly trained attention encoders are defined as,
\begin{eqnarray}
    \attnPoseW &=& E_P(\JposeW,\timeW) \nonumber \\
    \attnResidualPast &=& E_C(\JresidualWpast,\timeWpast), 
\end{eqnarray}
where $E_P$ and $E_C$ are multi-head attention networks, \JposeW and \JresidualWpast are the block of sequential Jacobians in the current window $W$ and the past window $W-1$, respectively; \timeW and \timeWpast are the corresponding blocks of time instances in these windows, which are positionally encoded.   

\begin{figure}[H]
    \centering
\includegraphics[width=\columnwidth]{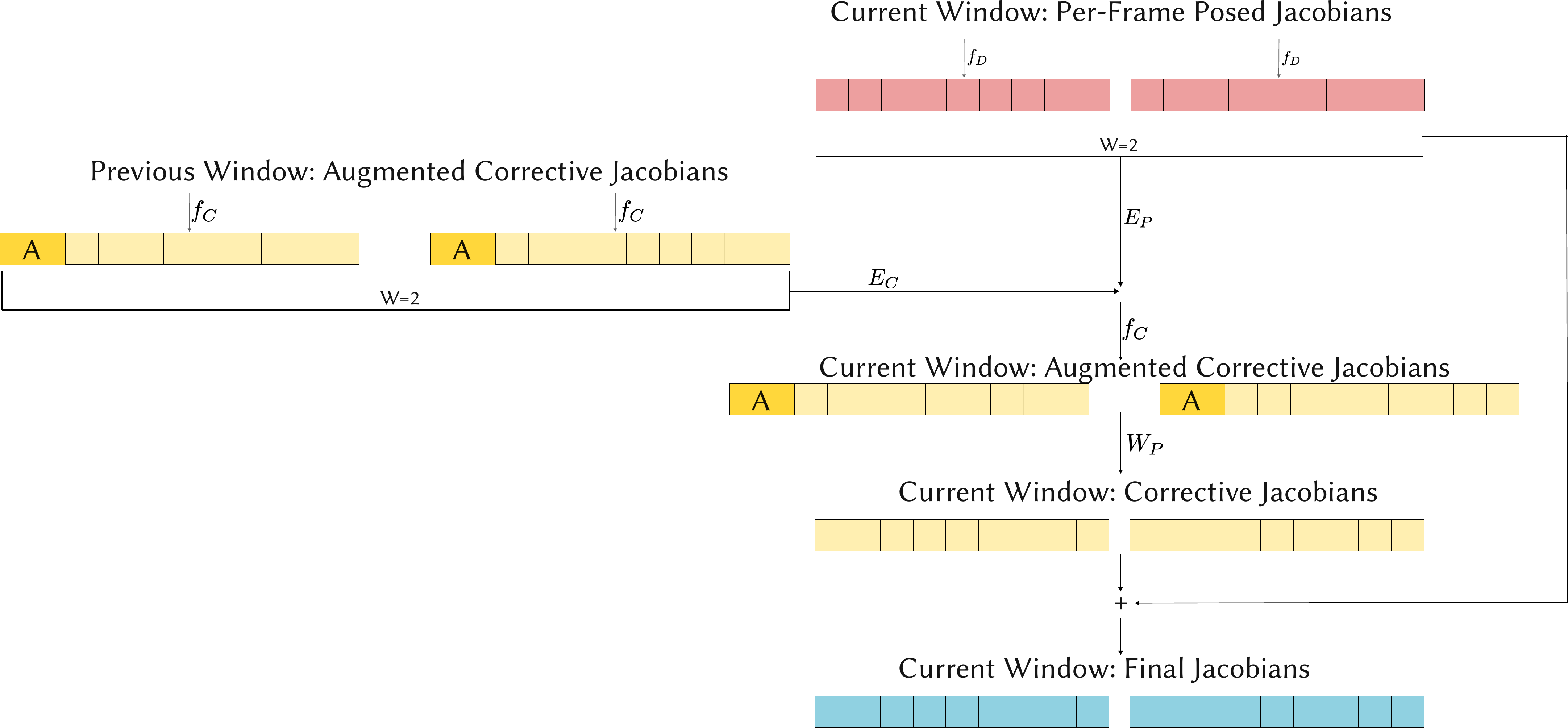}
\Description{Windowed Jacobian Prediction}
    \caption{
    {\textbf{Windowed Jacobian Prediction.} We use attention encodings of the current window's posed Jacobians (Eq \ref{eq:deformjac}) and the previous window's augmented corrective Jacobians (Eq \ref{eq:funcResidual}) to predict the current window's augmented corrective Jacobians. These are projected to predict the current window's corrective Jacobians (Eq \ref{eq:finalcorrective}). These corrective's are then added to the posed Jacobians to obtain the current window's final Jacobians. We use window size $W=32$.  } 
     }
\label{fig:window_jacobian_prediction}
\end{figure}

These attention networks encode a window of Jacobians into a single encoding as shown in \Cref{fig:window_jacobian_prediction}. Fixing the encoding sizes to a constant size enables handling any arbitrary window/sequence length without overflowing memory. The encoders distill the current window of Jacobian predictions \JposeW and previously predicted window of Corrective Jacobians \JresidualWpast. We pass the output of the attention networks as conditioning to \Cref{eq:funcResidual} to integrate and obtain the correctives in \Cref{eq:ode} in the augmented space, before projecting them in \Cref{eq:finalcorrective} to obtain the final correctives. We add the predicted correctives to the posed Jacobians $J_k$ from \Cref{eq:deformjac}. Finally, the predicted Jacobians \deformjac are spatially integrated using a differentiable Poisson solve~\cite{Nicolet2021Large}, in the coordinate frame of the first frame, to obtain the predicted shape $X_k$ at frame number $k$.

\paragraph*{Loss terms.} We train \textit{end-to-end} using only a shape loss over vertices of $X_k$ and a Jacobian loss. No extra annotation is required for supervision. Our final objective function is,
\begin{equation}{\label{eq:lvertex}}
    \loss_{\text{vertex}} = \| X_k - X_{k}^{GT} \|^2 \quad \text{and} \quad 
   \loss_{\text{Jacobian}} = \| J_k - J_{k}^{GT} \|^2,
\end{equation}
aggregated together into the total loss as
$ 
    \loss = \loss_\text{vertex} + \alpha \loss_\text{Jacobian}
$ with $\alpha=0.05$ in our experiments.

\begin{figure*}[h!]
    \centering
\includegraphics[width=0.99\linewidth]{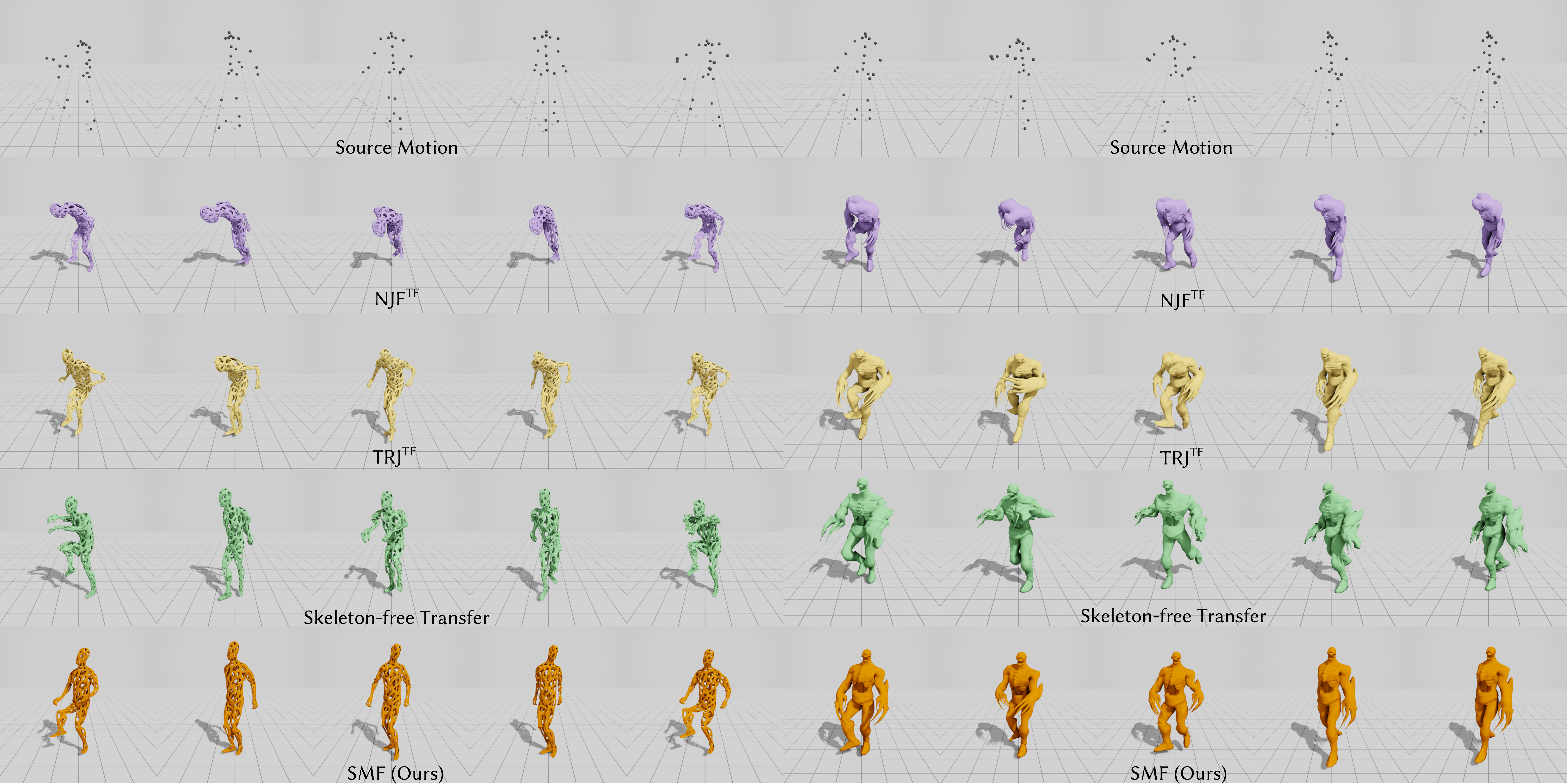}    
    \caption{
    {\textbf{Unseen motion from Out-of-Distribution dataset (Mixamo) applied to in-the-wild shapes.} We compare \shortname with NJF,  TRJ, and Skeleton-free transfer on  unseen dance motions (left: \texttt{Hiphop}; right: \texttt{Shuffle}) sampled from the out-of-distribution Mixamo dataset, applied to a 3D character found in-the-wild (hole man, left) and a Mixamo character (zombie, right). We modified NJF, TRJ to use keypoints instead, indicated by superscript \texttt{TF}. Competing methods exhibit distortion artifacts while attempting to follow the sampled source motion, while SMF (Ours) more accurately follows the sampled motion.
     }}
     \label{fig:unseen}
\end{figure*}

\begin{figure*}[h!]
    \centering
\includegraphics[width=0.98\linewidth]{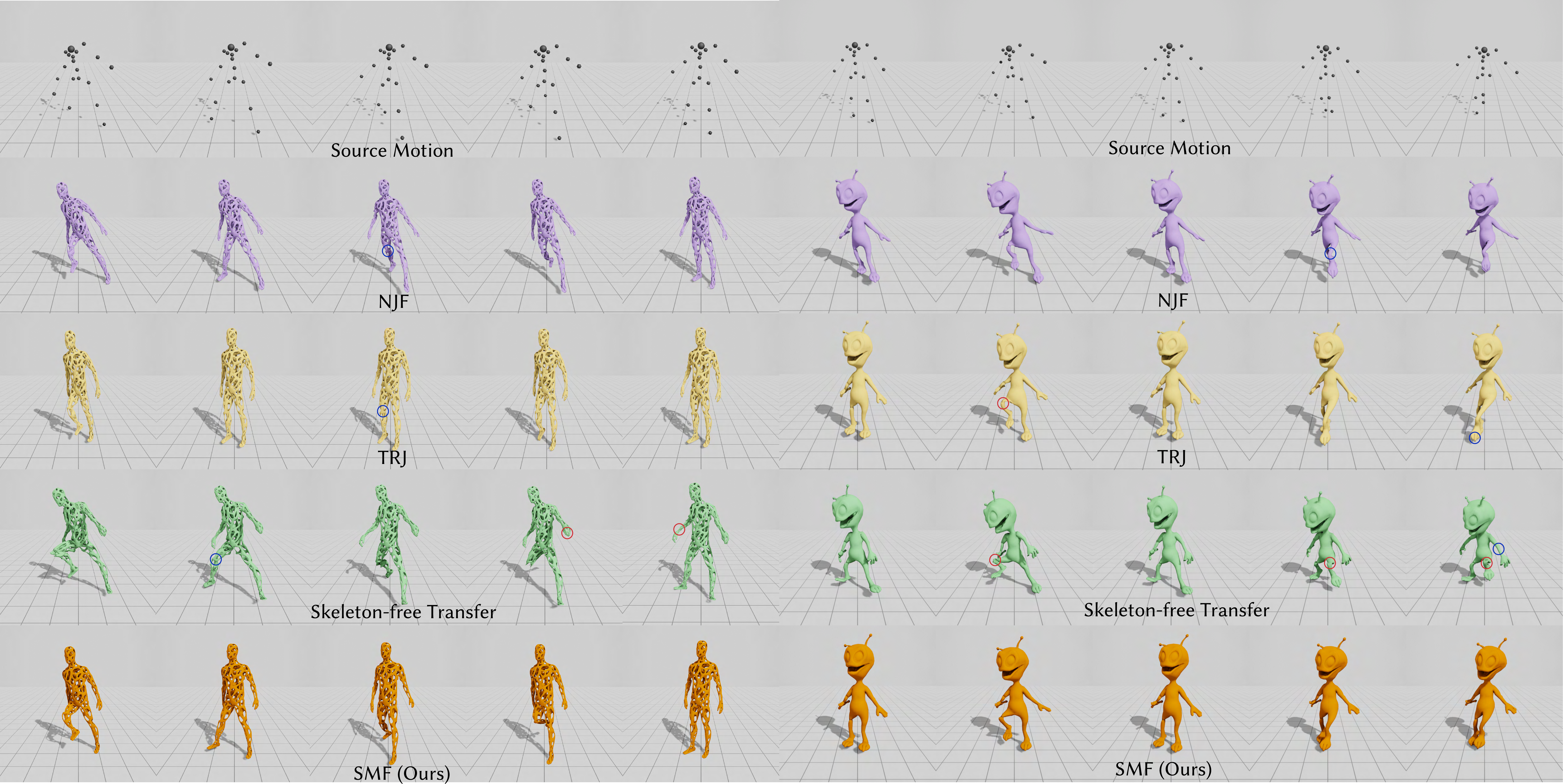}    
\Description{Comparison of motion transfer methods on 3D characters.}    
    \caption{
    {\textbf{Unseen motion applied to in-the-wild shapes.} We compare \shortname with NJF,  TRJ, and skeleton-free transfer on unseen motion (left: \texttt{Leg Backward Rotation}; right: \texttt{One Leg Jump}), applied to in-the-wild 3D characters. Baselines often  do not adhere to source motion (circled in blue) or exhibit distortion artifacts (circled in red). Our method transfers motion more accurately with far fewer shape distortion artifacts, while closely following the target motion.
     }}
     \label{fig:stylized}
\end{figure*}

\section{Evaluation}
\label{sec:results}

We test \name along multiple axes: (i)~generalization to unseen motion; (ii)~animation transfer to different characters (shapes and topology; meshes and scans); (iii)~motion specification using 3D keypoints, using monocular videos, or sampling/interpolating in the Kinetic Code (latent) space. No rigs or templates were used for any training or tests. We also test motion transfer from models trained on humanoid characters to non-humanoids. 

\setlength{\tabcolsep}{3pt}
\begin{table*}[t!]
\centering
\caption{
   \textbf{Quantitative evaluation unseen motion category, unseen shape.} We compare \name (ours) performance on unseen \textit{motion} categories, against multiple competing methods as well as ablated versions of our method.
We note that \name consistently produces results with the lowest errors, when compared against ground truth full body target meshes. Please refer to the supplemental webpage for videos. 
}
\label{tab:unseen-motion}
\footnotesize
\hspace{-1.9in}\resizebox{0.735\linewidth}{!}{
\begin{tabular*}{\linewidth}{@{\extracolsep{\fill}} 
r|cccc|cccc|cccc|cccc|cccc|cccc|cccc}
\multirow{2}{*}{\textbf{Method}} & \multicolumn{4}{c|}{\texttt{One Leg Jump}} & \multicolumn{4}{c|}{\texttt{Chicken Wings}} & \multicolumn{4}{c|}{\texttt{Walk}} & \multicolumn{4}{c|}{\texttt{Shake Shoulders}} & \multicolumn{4}{c|}{\texttt{Knees}} & \multicolumn{4}{c|}{\texttt{Shake Arms}} & \multicolumn{4}{c}{\texttt{Shake Hips}} \\
& L2-V & L2-$\delta$V & L2-J & L2-N & L2-V & L2-$\delta$V & L2-J & L2-N & L2-V & L2-$\delta$V & L2-J & L2-N & L2-V & L2-$\delta$V & L2-J & L2-N & L2-V & L2-$\delta$V & L2-J & L2-N & L2-V & L2-$\delta$V & L2-J & L2-N & L2-V & L2-$\delta$V & L2-J & L2-N\\
\cmidrule(lr){1-29}
Skeleton-free transfer \cite{skeletonfree} & 4.60 & 0.67 & 0.44 & 13.67 & \textbf{3.27} & 0.65 & 0.50 & 13.78 & 5.27 & 0.78 & 0.65 & 19.27 & 5.27 & 0.72 & 0.46 & 13.97 & 4.71 & 0.62 & 0.40 & 12.82 & 3.74 & 0.70 & 0.52 & \textbf{15.39} & 3.83 & 0.75 & 0.56 & 16.19 \\
\shortname (3D) w/o \inpmotion encoding  & 4.56 & 0.70 & 0.32 & 12.56 & 4.22 & 0.68 & 0.35 & 12.86 & 5.84 & 0.71 & 0.34 & 12.92 & 4.19 & 0.68 & 0.30 & 11.85 & 5.20 & 0.77 & 0.36 & 13.79 & 5.21 & 0.80& 0.44 & 19.79 & 4.48 & 0.71 & 0.35 & 13.26\\
\shortname (2D keypoints) & 5.74 & 0.72 & 0.33 & 12.42 & 6.50 & 0.80 & 0.44 & 19.43 & 6.87 & 0.79 & 0.36 & 13.49 & 6.77 & 0.87 & 0.38 & 15.63 & 8.33 & 0.88 & 0.39 & 14.81
 & 8.81 & 0.96 & 0.50 & 22.39 & 7.74 & 0.89 & 0.43 & 17.06\\
\shortname (3D keypoints) & \textbf{2.79} & \textbf{0.51} & \textbf{0.25} & \textbf{9.82} & 3.74 & \textbf{0.61} & \textbf{0.34} & \textbf{12.72} & \textbf{3.73} & \textbf{0.54} & \textbf{0.27} & \textbf{10.24} & \textbf{3.09} & \textbf{0.58} & \textbf{0.28} & \textbf{11.44} & \textbf{3.78} & \textbf{0.63} & \textbf{0.30} & \textbf{11.27} & \textbf{4.41} & \textbf{0.69} & \textbf{0.41} & 18.45 & \textbf{3.40} & \textbf{0.58} & \textbf{0.31} & \textbf{12.28}
\end{tabular*}
}
\end{table*}

\setlength{\tabcolsep}{8pt}
\begin{table*}[t!]
\centering
\caption{
   \textbf{Quantitative evaluation on Mixamo motion (unseen dataset), unseen stylized character.} We compare \name (ours) against competing methods and report averaged errors (over 3 characters per motion) against rig-based Mixamo~\cite{Mixamo} meshes as groundtruth. To work with Mixamo, we modified TRJ and NJF to use keypoints instead of SMPL parameters, indicated by the superscript \texttt{TF}. %
}
\label{tab:unseen-motion-mixamo}
\footnotesize
\hspace{-2.1in}\resizebox{0.69\linewidth}{!}{
\begin{tabular*}{\linewidth}{@{\extracolsep{\fill}} r|cccc|cccc|cccc|cccc|cccc}
\multirow{2}{*}{\textbf{Method}} & \multicolumn{4}{c|}{\texttt{Hiphop}} & \multicolumn{4}{c|}{\texttt{Shuffling}} & \multicolumn{4}{c|}{\texttt{Surprised}} & \multicolumn{4}{c|}{\texttt{Shaking Hands}} & \multicolumn{4}{c}{\texttt{Arguing}} \\
& L2-V & L2-$\delta$V & L2-J & L2-N & L2-V & L2-$\delta$V & L2-J & L2-N & L2-V & L2-$\delta$V & L2-J & L2-N & L2-V & L2-$\delta$V & L2-J & L2-N & L2-V & L2-$\delta$V & L2-J & L2-N\\
\cmidrule(lr){1-21}
Skeleton-free transfer \cite{skeletonfree} & 12.41& 4.61& 1.08& 32.28 & 13.72& 3.81 & 0.83& 25.51 & 13.24& 2.79& 0.70& 21.14 & 17.44& 2.75& 0.67& 20.48 & 14.35 & 2.60 & 0.69 & 20.85 \\
$NJF^{TF}$ \cite{aigerman2022neural} & 25.57& 9.03& 1.13& 45.94 & 25.38& 9.02& 1.11& 45.67 & 26.27& 8.96& 1.10& 45.24 & 31.80& 10.68& 1.19& 45.80 & 29.22& 9.71& 1.14& 45.26 \\
$TRJ^{TF}$ \cite{trj} & 15.23& 5.86& 0.87& 35.09 & 16.91& 6.62& 0.85& 33.54 & 14.22& 5.20& 0.64& 24.95 & 15.99& 5.94& 0.73& 30.25 & 15.17& 5.39& 0.70& 28.62 \\
\shortname (3D keypoints) & \textbf{8.35} & \textbf{3.58} & \textbf{0.62} & \textbf{24.38} & \textbf{7.97} & \textbf{3.50} & \textbf{0.60} & \textbf{23.46} & \textbf{4.60} & \textbf{2.00} & \textbf{0.28} & \textbf{10.36} & \textbf{5.78} & \textbf{2.31} & \textbf{0.43} & \textbf{17.86} & \textbf{5.98} & \textbf{2.27} & \textbf{0.40} & \textbf{16.03} \\

\end{tabular*}
}
\end{table*}

\begin{table*}[h]
    \centering
    \caption{\textbf{Generalization of Kinetic Codes.} We compare the reconstruction error of the kinetic codes of seen and unseen motion. The minimal variation in reconstruction errors indicates the generalizability of our codes to varied unseen motions.}
    \resizebox{\linewidth}{!}{\begin{tabular}{lcccc|ccccccc}
        & \multicolumn{4}{c}{Seen Motion} & \multicolumn{7}{c}{Unseen Motion} \\
        \cmidrule(lr){2-5} \cmidrule(lr){6-12}
        & \texttt{Running} & \texttt{Jumping Jacks} & \texttt{Punching} & \texttt{Jiggle on Toes} 
        & \texttt{Shake Hips} & \texttt{Hips} & \texttt{Shake Shoulders} & \texttt{One Leg Jump} & \texttt{Shake Arms} & \texttt{Walk} & 
        \texttt{Chicken Wings} \\
        \midrule
        Error (in $10^{-3}$ cm) & 7.73 & 11.88 & 12.03 & 11.13 
              & 6.88 & 8.36 & 8.34 & 8.51 & 7.95 & 8.96 & 7.89 \\
    \end{tabular}}
\end{table*}

\paragraph{Motion datasets.}
We train our method (and baselines) on a single dataset comprising of $5$ human motion categories from the AMASS dataset~\cite{AMASS:ICCV:2019} with each category containing approximately 6-7 motion sequences. We evaluate our method on unseen motion categories from AMASS, sampled motions from Mixamo, animal motion sequences from DeformingThings4D dataset~\cite{def4d}, and in-the-wild monocular video recordings. The AMASS dataset utilizes the SMPL body model~\cite{SMPL:2015}, which enables generation of motion sequences for diverse body-shapes by varying the shape parameter, $\beta$. Note that ours does \textit{not} use this SMPL information during training or inference.

To show that our method works on animals, we train our method on motions from the DeformingThings4D dataset~\cite{def4d}, which provides animal 4D meshes as deforming sequences and test it on unseen motion sequences.

We train \name on 9 motion categories from AMASS, each consisting of sequences performed by 6 humans. The number of frames varies from 150-800. For 2D keypoints, we use 3 YouTube videos (2 characters in total), totaling 1hr. Although the videos are noisy as the camera angles change, \name successfully learns due to the Kinetic codes setup.

\subsection{Baselines}\label{baselines}
We compare our method against recent per frame (pose transfer) methods:   NJF~\cite{aigerman2022neural} %
and Skeleton-Free Pose Transfer~\cite{skeletonfree}. Due to unavailability of pre-processing code for Skeleton-Free Pose Transfer, we use their pretrained models which were trained on substantially more data including AMASS and Mixamo motions as well as stylized characters. Additionally, it requires the full mesh sequence along with T pose meshes for source and target characters. In comparison, ours only takes in sparse keypoints as input.  We also compare ours with animation transfer methods: TRJ~\cite{trj} (which uses template) and an ablated version of our method without the Kinetic Code (\inpmotion) encoding. In the table below, we highlight the differences between \name and baseline methods. 

\begin{table}[H]
\centering
\resizebox{\linewidth}{!}{%
\begin{tabular}{r c c c c}
\textbf{Methods} & \textbf{Rig-Free} & \textbf{Template-Free} & \textbf{Temporally-Coherent} & \textbf{Self-Supervised} \\
\toprule
NJF~\cite{aigerman2022neural}& $\ding{51}$ & $\ding{51}$ & \ding{55} & \ding{55} \\
Skeleton Free~\cite{skeletonfree} & \ding{55} & $\ding{51}$ & \ding{55} & \ding{55} \\
TRJ~\cite{trj} & $\ding{51}$ & \ding{55} & $\ding{51}$ & \ding{55} \\
SMF (ours) & $\ding{51}$ & $\ding{51}$ & $\ding{51}$ & $\ding{51}$ \\
\bottomrule
\end{tabular}%
}
\centering
\resizebox{\linewidth}{!}{%
\begin{tabular}{r l}
\textbf{Method} & \textbf{Input Requirements} \\
\toprule
NJF~\cite{aigerman2022neural} & SMPL Pose Parameters \\
Skeleton Free~\cite{skeletonfree} & T-Posed Source \& Target characters, full body source motion \\
TRJ~\cite{trj} & SMPL Pose and Bodyshape Parameters\\
SMF (ours) & Sparse Keypoints \\
$NJF^{TF}$ & Modified \& trained with Keypoints \\
$TRJ^{TF}$ & Modified \& trained with Keypoints \\
\bottomrule
\end{tabular}%
}
\label{tab:methods_input_comparison}
\end{table}

\subsection{Metrics and Target Shapes}\label{metrics}
We evaluate our animation transfer with four main metrics:
\begin{itemize}
    \item \textbf{Vertex-to-vertex error (L2-V)}: Measures the Euclidean distance between ground-truth and predicted mesh vertices, indicating how well the global motion is captured.
    \item \textbf{Velocity error (L2-\(\delta\)V)}: Quantifies differences in vertex velocities across frames, capturing the temporal smoothness and cohesion of the animation.
    \item \textbf{Jacobian error (L2-J)}: Assesses deviations in local transformations, revealing unintended deformations.
    \item \textbf{Angular error of surface normals (L2-N)}: Calculates the angle between predicted and ground-truth normals, indicating preservation of local surface orientations.
\end{itemize}

\paragraph{Target Shapes}
We evaluate our method on diverse target shapes varied body shapes sampled from SMPL models, human scans from the FAUST dataset \cite{bogo2014faust}, characters from the Mixamo library (e.g., skeleton zombie, triceratop, wolf), and in-the-wild meshes from online 3D repositories (e.g., alien, hole-man). As preprocessing, when applicable, we fixed non-manifold meshes.

\subsection{Qualitative Results} 
\paragraph*{Generalization to unseen motion and shape.} We present video results on various unseen motion and unseen shapes on our supplementary webpage. Our method produces consistently better generalization to unseen motion categories compared to NJF and TRJ, both of which result in shape distortion and erroneous displacements as they struggle to follow the input (unseen) motion as seen in \Cref{fig:unseen}. 
This is particularly highlighted in scenarios with large displacements, e.g., feet and hands are widened or stretched thin. 
Note that unlike in the original TRJ~\cite{trj}, where specialized  models were separately trained for each motion type, we retrained a single TRJ, across all the motion types.

\name also generalizes to new shapes of varied body types, including non-humans despite being trained only on humans. This correspondence from coarse keypoints to a dense mapping across diverse shapes is learned during the self-supervised motion transfer setup and proves to be even capable of generalizing to multi-legged creatures. It preserves the source motion and target shape and the resultant motion is realistic and free from jitters/artifacts. We show comparative results on unseen motion transfer to in-the-wild target meshes in \Cref{fig:unseen}. For high-genus shapes, such as the mesh with holes (right half of \Cref{fig:unseen}), NJF distorts the shape; TRJ fares comparatively better, it is still riddled with artifacts. Our Kinetic codes not only preserve the target shape but also more faithfully adhere to the motion. See \href{https://motionfields.github.io/}{\revision{supplemental webpage}} for video results. %

\begin{figure}[htbp]
    \centering
\includegraphics[width=0.945\columnwidth]{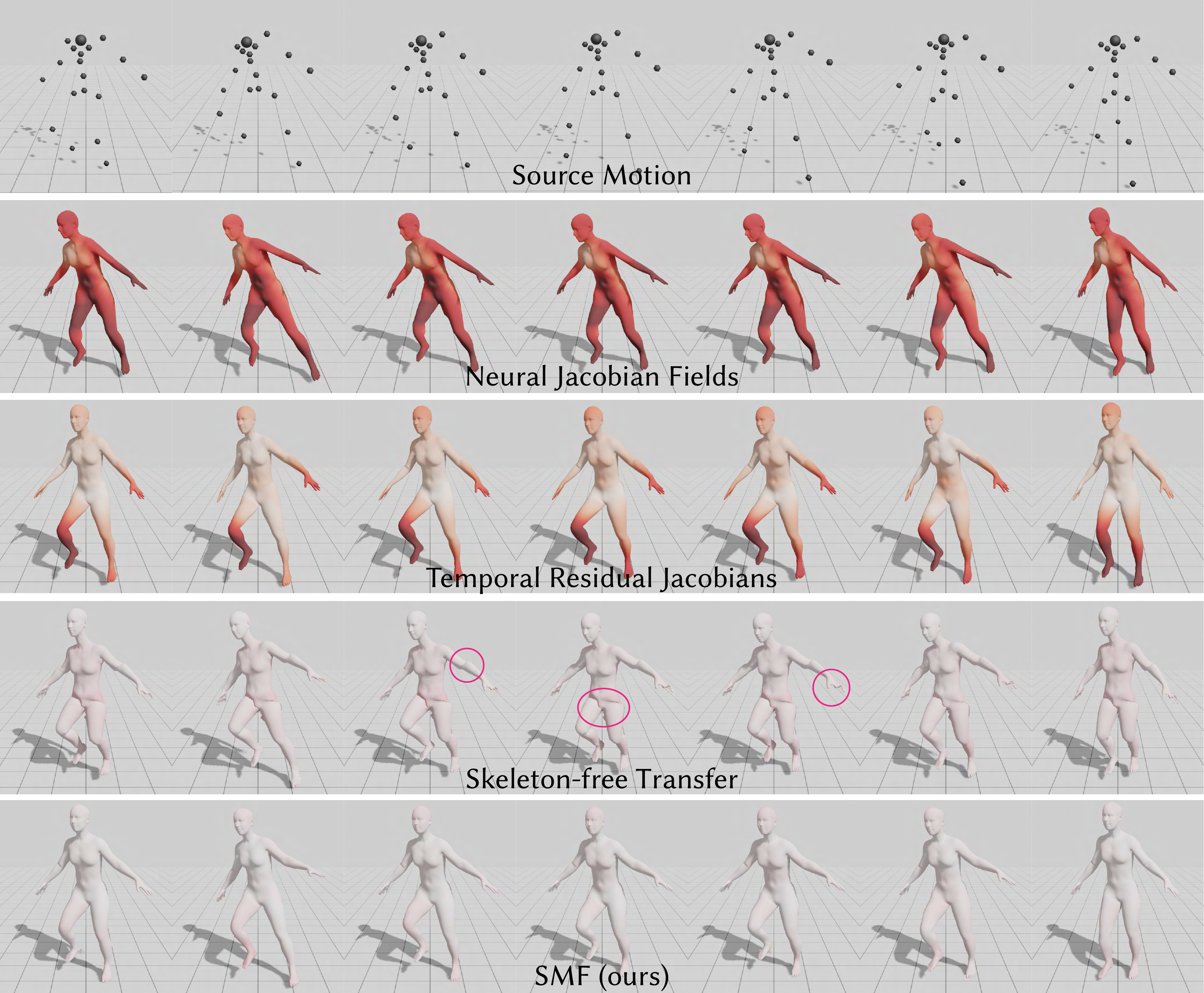}
    \Description{Ablations.}
    \caption
    {\textbf{Comparison of \shortname with baselines.} We compare \shortname with Neural Jacobian Fields~\cite{aigerman2022neural}, Temporal Residual Jacobians~\cite{trj}, and template-free skeleton-free transfer~\cite{skeletonfree}. We measure the vertex-to-vertex error with ground truth and color-code the results according to the measured error. Darker red indicates higher error. \shortname accurately transfers the motion to the target mesh, while baselines struggle to follow the input motion and exhibit distortion artifacts.}
    \label{fig:ablation}
\end{figure}

Utilizing a latent representation for motion encoding with smooth interpolation properties leads to improved generalization. Hence, we do not see any significant artifacts even when operating on in-the-wild unseen target shapes for unseen motion. Moreover, we notice without our motion encoding, regions around joints may show melting (see around feet in \Cref{fig:ablation}). The same holds true for more accurate bending of the joints (see \Cref{fig:unseen}). We note that the AMASS motions contain foot-skating artifacts, and ours faithfully reproduces them. However, we also tested ours on high-quality Mixamo motions as shown in \Cref{fig:unseen} and supplemental webpage, where foot skating is less pronounced. Despite being trained only on AMASS, our model successfully transfers these clean Mixamo motions without introducing any additional skating artifacts. This demonstrates the model's generalization not only to unseen motions but also to different levels of motion quality. This further suggests that our self-supervised setting is promising  and training on higher-quality motion samples would likely yield better results.

\begin{figure}[htbp]
    \centering
\includegraphics[width=0.945\columnwidth]{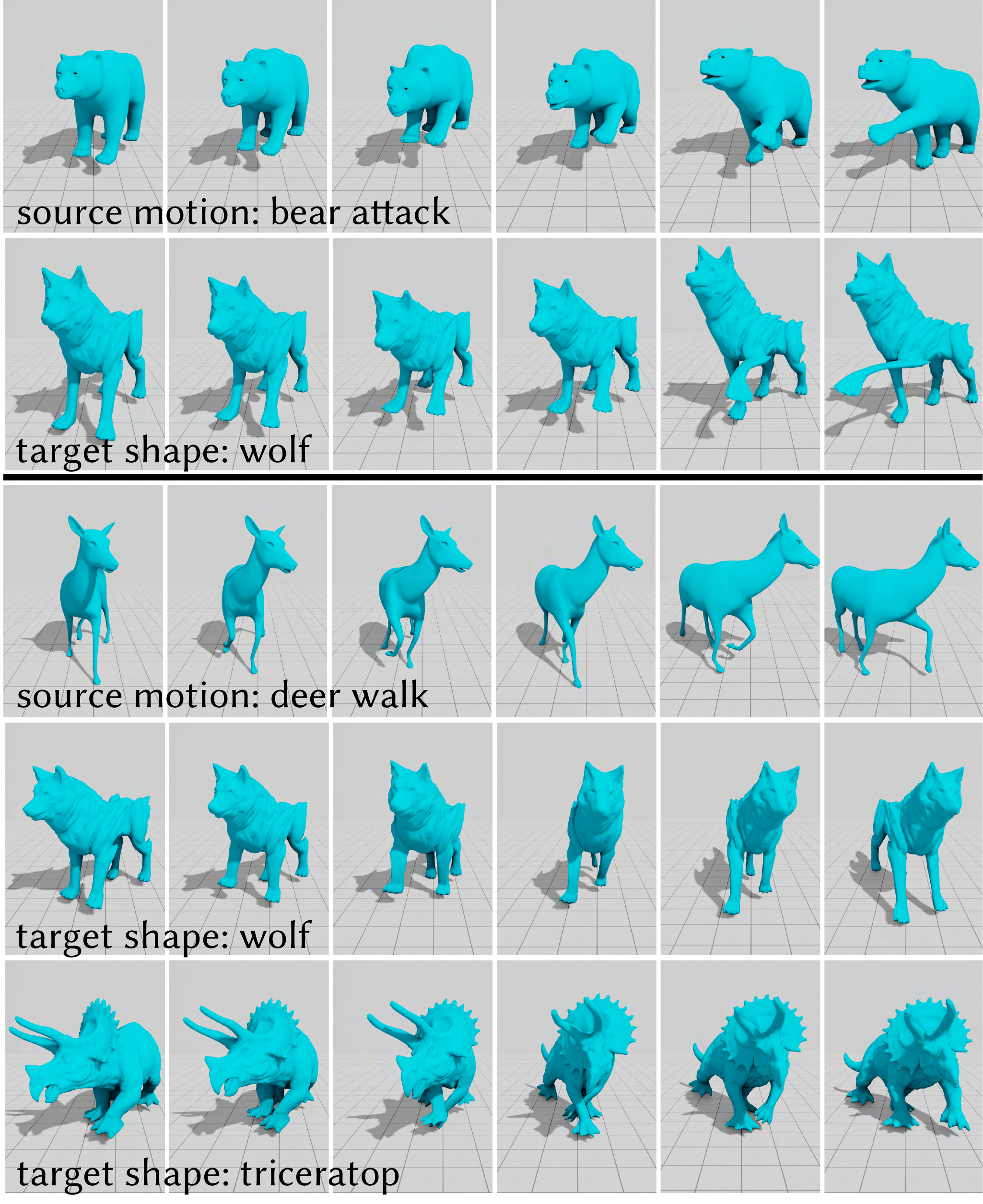}
\Description{Results on D4D}
    \caption{
    {\textbf{Unseen motion to animal shapes.} We use \shortname for animation transfer to animal shapes (wolf and triceratop) for different source motions (top: \texttt{bear attack}; bottom: \texttt{deer walk}). Our method transfers motion faithfully while closely following the target motion (see supplemental videos).
     }}
\label{fig:animals_comparison}
\end{figure}

\begin{figure*}[t!]
    \centering
\includegraphics[width=\linewidth]{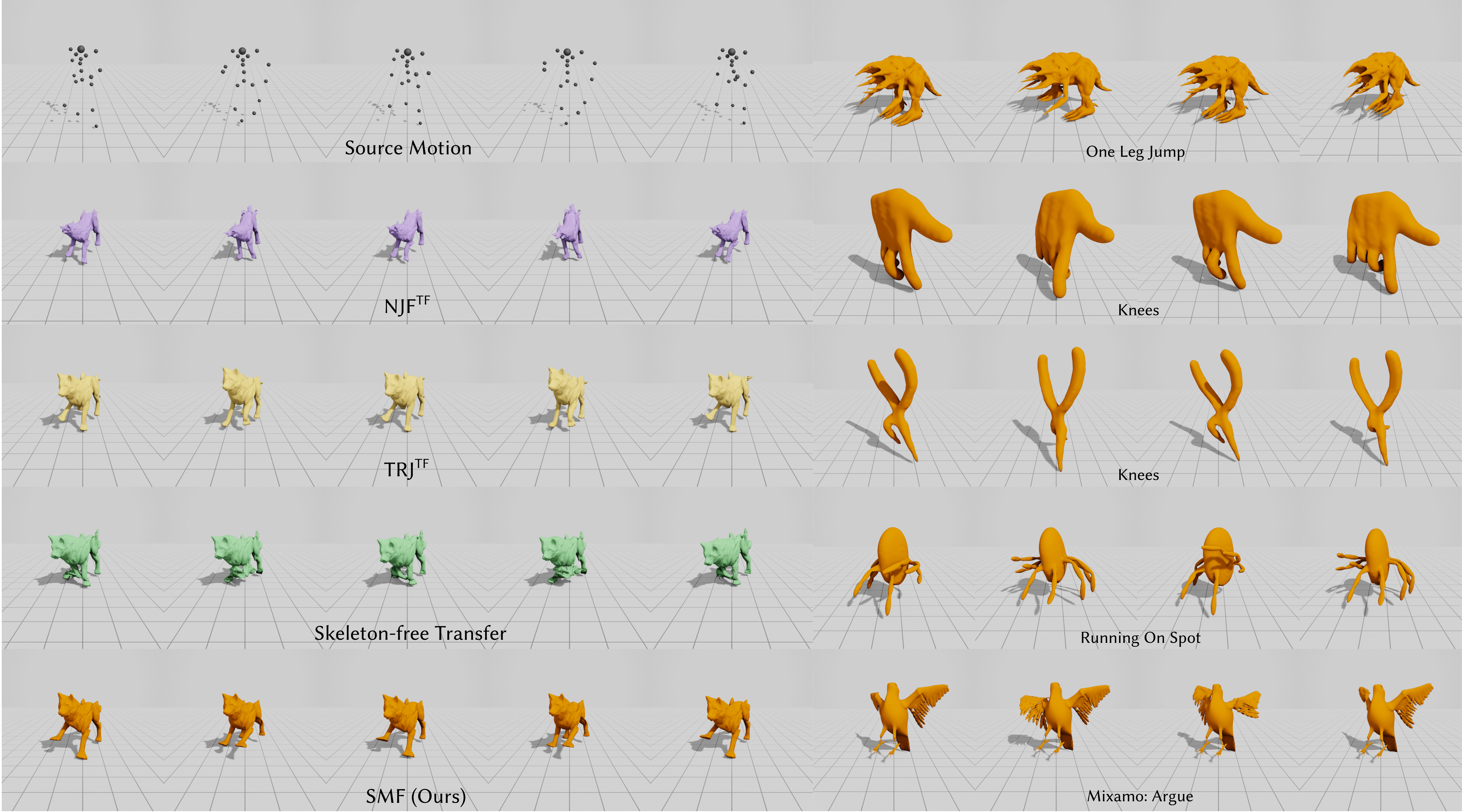}
    \caption
    {\textbf{Human to non-human motion transfer} \textit{Left:} We compare \shortname with 
    NJF, TRJ, and skeleton-free transfer on the task of transferring human motion to non-human characters. Since bodyshape parameters are unavailable for these animals, we utilize modified NJF and TRJ models trained on keypoints, as indicated by superscript \texttt{TF}. \shortname, without using SMPL bodyshape parameters, generalizes to wider range of shapes without distortion artifacts. \textit{Right:} \shortname transfers varied unseen motions to several Out-of-Distribution characters. Unlike template-based methods, \shortname does not use any body-shape descriptor.
     }
\label{fig:nonhuman}
\end{figure*}

\Cref{fig:teaser} and \Cref{fig:nonhuman} show more animation transfer examples to different humanoids and non-humanoids target meshes,  sourced from online in-the-wild meshes and the SHREC'07 dataset~\cite{giorgi2007shape}. Moreover, \Cref{fig:teaser} showcases support for different types of coarse motion specifications. 
In \Cref{fig:animals_comparison}, we show  retargetting results on target animal shapes from the D4D dataset.

\paragraph*{Generalization to Mixamo sampled motions on stylized characters.} To transfer motion from Mixamo~\cite{Mixamo}, we map its joint locations (3D keypoints) to our system, (note the mapping is not perfect due to misalignment between joint locations, requiring some approximation). Our results demonstrate strong generalization, transferring these motions even to different characters. Notably, TRJ~\cite{trj} and NJF~\cite{aigerman2022neural} are unable to operate on Mixamo motion, as they strictly depend on the SMPL template, which significantly limits their flexibility. Therefore, we modify their architecture to handle 3D coordinates and remove the shape parameter ($\beta$) module from TRJ.  Please see supplemental videos on the webpage for comparison. We further evaluate it on 5 different motions on 3 stylized characters with ground truth generated from Mixamo. As seen in \Cref{tab:unseen-motion-mixamo}, we see large improvements over existing methods. Except Skeleton-free transfer (additionally trained on Mixamo data and stylized characters), all other baselines have been trained solely on the AMASS dataset with human characters. 

\begin{figure*}[t!]
    \centering

    \includegraphics[width=\linewidth]{figures/dance_comparison.pdf}\\[4mm]

    \includegraphics[width=0.32\textwidth]{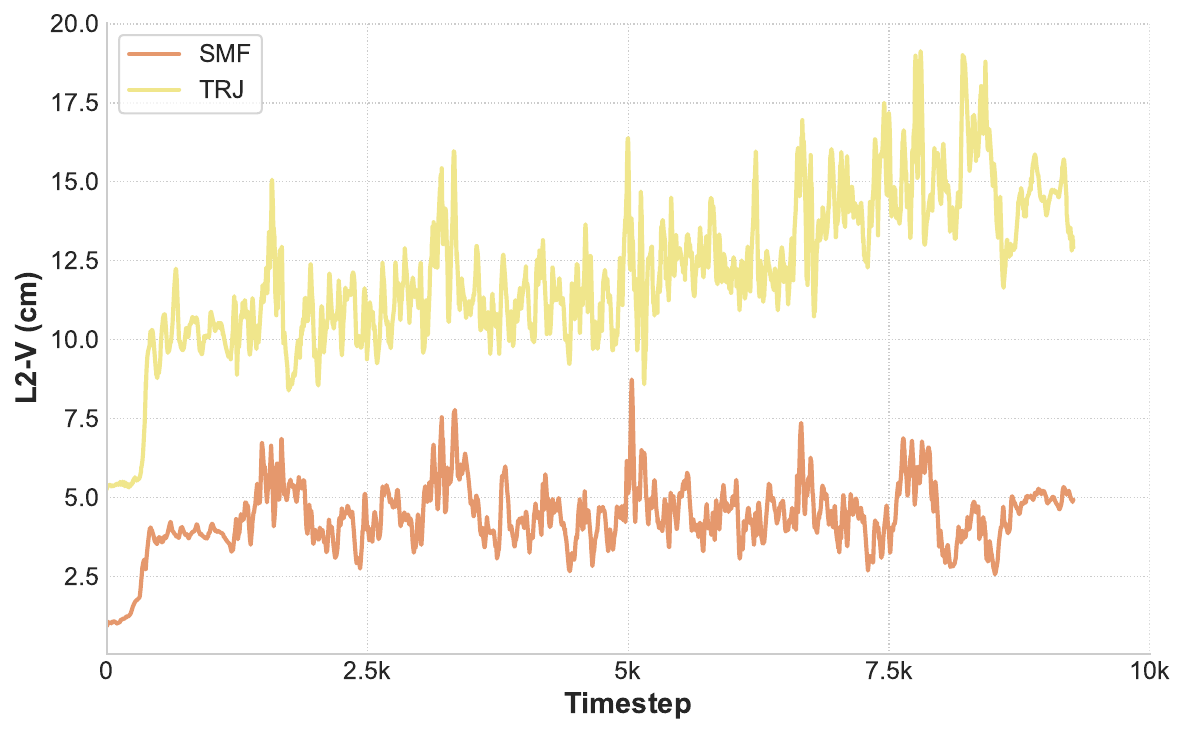}\hfill%
    \includegraphics[width=0.32\textwidth]{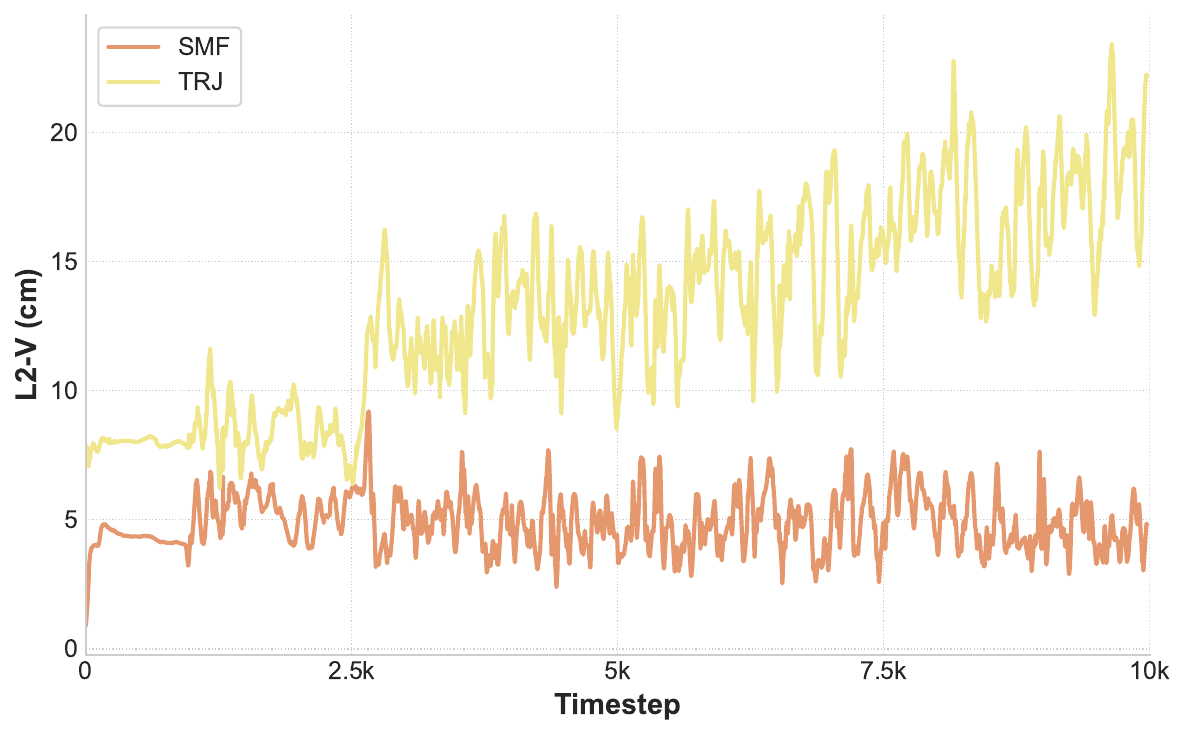}\hfill%
    \includegraphics[width=0.32\textwidth]{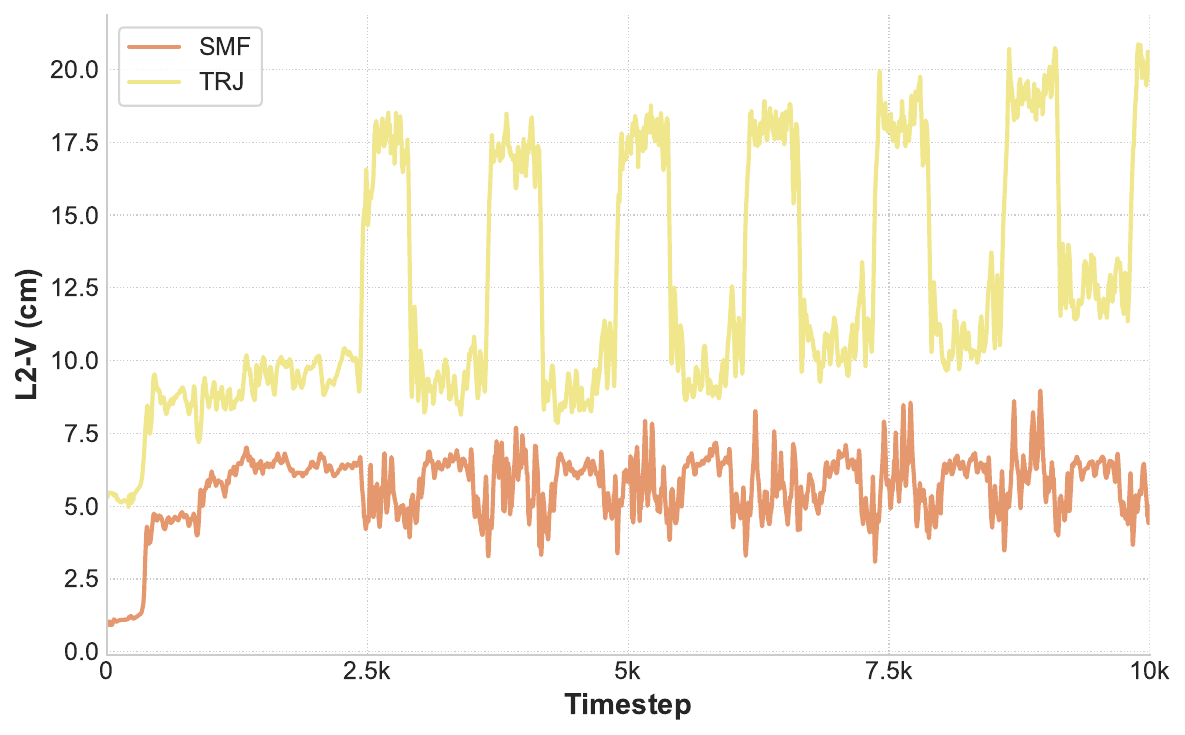}
    
    \caption{
    {\textbf{Temporal stability and comparison on very long motion sequences.}
    We compare \shortname and TRJ on various \textit{unseen} folk dance sequences (10000 frames) from the AMASS dataset.
    \textbf{Top.} Qualitative comparison on two unseen dance sequence. As the sequence progresses (frame numbers at top), TRJ accumulates significant error, exacerbating distortion artifacts. In contrast, \shortname follows the source motion more closely.
    \textbf{Bottom.} Per-frame vertex-to-vertex (L2-V) error plots for three unseen dance sequences. TRJ displays a monotonically increasing error trend due to accumulation, while \shortname maintains a consistently low error profile.
    This demonstrates \shortname's robustness to temporal drift and its suitability for generating long motion sequences.
    }}
    \label{fig:long_motion_comparison}
\end{figure*}

\paragraph*{Generalization to monocular capture.} \shortname can directly transfer raw motion sequences from 2D input videos captured on handheld cameras to a 3D target shape. We use our formulation based on 2D keypoints to directly operate on 2D input frames. This is a particularly challenging problem as we are lifting motion from 2D to a 3D shape \textit{without} any explicit supervision or template shape. For our 2D training setup, we additionally train only the autoencoder on in-the-wild videos of exercise videos totaling 1 hour. We note this is noisy but this helps the \inpmotion generalize to RGB captures better as all our 2D renderings of AMASS sequences have a similar camera setup. We extract 2D keypoints using HRNet~\cite{hrnet}, which computes 2D keypoints frame by frame. We use Savitzky-Golay filter to smoothen the 2D signal as there is inter-frame noise. 

\noindent\paragraph*{Self-supervised training.} Our self-supervised setup enables us to learn motion transfer without requiring a template shape and using only coarse motion parameters extracted at joints from the motion dataset. This setup is the key to enabling learning from coarse input motion (2D/3D keypoints) and transferring motion from 2D monocular captures and Mixamo sampled motion to 3D target shapes.

\subsection{Quantitative Comparison}
We present quantitative results on unseen motion category and unseen motion sequence (within the same category) to unseen shape in \Cref{tab:unseen-motion,tab:unseen-motion-mixamo}. Generalization to unseen motion is a particularly challenging problem as the Poisson solve in NJF and TRJ facilitates shape preservation for unseen shape by following the coordinates of the source shape. However, there is no such aid when generalizing to unseen motion, which is an extrapolation problem. Our method significantly outperforms NJF and TRJ for both motion accuracy as well as detail preservation. The benefits of our \inpmotion are highlighted here as we see significantly lower errors against a baseline of \shortname without \inpmotion encoding. Compared to the per-frame pose transfer performed by Skeleton-Free Pose Transfer which has numerous artifacts and unnatural deformations, SMF also ensures temporal consistency as seen in lower velocity error (L2-$\delta$V), measuring motion smoothness.

\paragraph{Error accumulation on long motion transfer.} We choose 3 unseen, complex, and long dance sequences (10000 frames) to analyze error accumulation. We report per frame vertex-to-vertex (L2-V) error in \Cref{fig:long_motion_comparison}. Compared to the only other temporal method, TRJ, SMF does not exhibit a trend of monotonically increasing error demonstrating its robustness to temporal drift and its suitability for generating long motion sequences. The initial spike in error is due to the motion moving away from the easy-to-model rest pose.

\subsection{Performance}
\name's lightweight architecture enables inference at >30 FPS on a single RTX 4090 for a mesh with ~30k faces (for high number of faces this will be slower). This makes our system suitable for real-time applications, following a one-time preprocessing step to compute features on the target mesh. Furthermore, when using 2D keypoints, the system could potentially be extended for real-time 2D-to-3D animation transfer, as keypoint extraction with HRNet~\cite{hrnet} is highly efficient.

\section{Conclusion}

We have presented \textit{Self-supervised Motion Fields} that convert sparse motion into full-body character motion. We enable this by first creating a temporally-informed motion latent space -- Kinetic codes --  and then utilizing it to train spatial and temporal gradient predictor networks jointly. The gradient signals are coupled via spatial and temporal integration, and trained using full-body mesh sequences for supervision. \name does \textit{not} require additional motion annotation, is simple and robust to train, and generalizes across unseen motion and character shape variations. We also do not use any template model or rig, either during training or inference. We establish a new SOTA on the AMASS dataset. 

\textit{Limitations.} While our formulation captures some secondary motion, we do not model secondary dynamics in any physically correct way. Neither do we explicitly support collision handling to prevent self-intersection (see the triceratop leg crossing in \Cref{fig:animals_comparison}). An interesting future direction would be to implicitly handle collision detection and simultaneously model character-garment interactions, possibly using a transformer-based attention mechanism to capture non-local interactions between body parts. Similar to single human bodies~\cite{Fruehstueck2022InsetGAN}, it will be interesting to combine different motion generators for different parts of a character, allowing mixing of motion modes. Finally, temporal sketching sequences~\cite{fzlm_animatedLineDraw_siga_11} can also be modeled with such a generative framework.

\clearpage
\bibliographystyle{ACM-Reference-Format}
\bibliography{main}


\begin{thebibliography}{47}

%%% the Web address.

\ifx \showCODEN    \undefined \def \showCODEN     #1{\unskip}     \fi
\ifx \showISBNx    \undefined \def \showISBNx     #1{\unskip}     \fi
\ifx \showISBNxiii \undefined \def \showISBNxiii  #1{\unskip}     \fi
\ifx \showISSN     \undefined \def \showISSN      #1{\unskip}     \fi
\ifx \showLCCN     \undefined \def \showLCCN      #1{\unskip}     \fi
\ifx \shownote     \undefined \def \shownote      #1{#1}          \fi
\ifx \showarticletitle \undefined \def \showarticletitle #1{#1}   \fi
\ifx \showURL      \undefined \def \showURL       {\relax}        \fi
% invisible to TeX
\providecommand\bibfield[2]{#2}
\providecommand\bibinfo[2]{#2}
\providecommand\natexlab[1]{#1}
\providecommand\showeprint[2][]{arXiv:#2}

\bibitem[Abdrashitov et~al\mbox{.}(2021)]%
        {abdrashitov2021musculoskeletal}
\bibfield{author}{\bibinfo{person}{Rinat Abdrashitov}, \bibinfo{person}{Seungbae Bang}, \bibinfo{person}{David Levin}, \bibinfo{person}{Karan Singh}, {and} \bibinfo{person}{Alec Jacobson}.} \bibinfo{year}{2021}\natexlab{}.
\newblock \showarticletitle{Interactive Modelling of Volumetric Musculoskeletal Anatomy}.
\newblock \bibinfo{journal}{\emph{ACM Transactions on Graphics}} \bibinfo{volume}{40}, \bibinfo{number}{4} (\bibinfo{year}{2021}).
\newblock


\bibitem[Aberman et~al\mbox{.}(2020)]%
        {skeleton-aware2020}
\bibfield{author}{\bibinfo{person}{Kfir Aberman}, \bibinfo{person}{Peizhuo Li}, \bibinfo{person}{Dani Lischinski}, \bibinfo{person}{Olga Sorkine-Hornung}, \bibinfo{person}{Daniel Cohen-Or}, {and} \bibinfo{person}{Baoquan Chen}.} \bibinfo{year}{2020}\natexlab{}.
\newblock \showarticletitle{Skeleton-aware networks for deep motion retargeting}.
\newblock \bibinfo{journal}{\emph{ACM Transactions on Graphics (TOG)}} \bibinfo{volume}{39}, \bibinfo{number}{4} (\bibinfo{year}{2020}), \bibinfo{pages}{62--1}.
\newblock


\bibitem[Aigerman et~al\mbox{.}(2022)]%
        {aigerman2022neural}
\bibfield{author}{\bibinfo{person}{Noam Aigerman}, \bibinfo{person}{Kunal Gupta}, \bibinfo{person}{Vladimir~G Kim}, \bibinfo{person}{Siddhartha Chaudhuri}, \bibinfo{person}{Jun Saito}, {and} \bibinfo{person}{Thibault Groueix}.} \bibinfo{year}{2022}\natexlab{}.
\newblock \showarticletitle{Neural Jacobian Fields: Learning Intrinsic Mappings of Arbitrary Meshes}.
\newblock \bibinfo{journal}{\emph{SIGGRAPH}} (\bibinfo{year}{2022}).
\newblock


\bibitem[Aubry et~al\mbox{.}(2011)]%
        {aubry2011wave}
\bibfield{author}{\bibinfo{person}{Mathieu Aubry}, \bibinfo{person}{Ulrich Schlickewei}, {and} \bibinfo{person}{Daniel Cremers}.} \bibinfo{year}{2011}\natexlab{}.
\newblock \showarticletitle{The wave kernel signature: A quantum mechanical approach to shape analysis}. In \bibinfo{booktitle}{\emph{2011 IEEE international conference on computer vision workshops (ICCV workshops)}}. IEEE, \bibinfo{pages}{1626--1633}.
\newblock


\bibitem[Baran and Popovi\'{c}(2007)]%
        {baran2007automatic}
\bibfield{author}{\bibinfo{person}{Ilya Baran} {and} \bibinfo{person}{Jovan Popovi\'{c}}.} \bibinfo{year}{2007}\natexlab{}.
\newblock \showarticletitle{Automatic rigging and animation of 3D characters}.
\newblock \bibinfo{journal}{\emph{ACM Trans. Graph.}} \bibinfo{volume}{26}, \bibinfo{number}{3} (\bibinfo{year}{2007}), \bibinfo{pages}{72–es}.
\newblock


\bibitem[Bogo et~al\mbox{.}(2014)]%
        {bogo2014faust}
\bibfield{author}{\bibinfo{person}{Federica Bogo}, \bibinfo{person}{Javier Romero}, \bibinfo{person}{Matthew Loper}, {and} \bibinfo{person}{Michael~J Black}.} \bibinfo{year}{2014}\natexlab{}.
\newblock \showarticletitle{FAUST: Dataset and evaluation for 3D mesh registration}. In \bibinfo{booktitle}{\emph{Proceedings of the IEEE conference on computer vision and pattern recognition}}. \bibinfo{pages}{3794--3801}.
\newblock


\bibitem[Chadwick et~al\mbox{.}(1989)]%
        {chadwick1989layered}
\bibfield{author}{\bibinfo{person}{John~E Chadwick}, \bibinfo{person}{David~R Haumann}, {and} \bibinfo{person}{Richard~E Parent}.} \bibinfo{year}{1989}\natexlab{}.
\newblock \showarticletitle{Layered construction for deformable animated characters}.
\newblock \bibinfo{journal}{\emph{ACM Siggraph Computer Graphics}} \bibinfo{volume}{23}, \bibinfo{number}{3} (\bibinfo{year}{1989}), \bibinfo{pages}{243--252}.
\newblock


\bibitem[Dupont et~al\mbox{.}(2019)]%
        {dupont2019augmented}
\bibfield{author}{\bibinfo{person}{Emilien Dupont}, \bibinfo{person}{Arnaud Doucet}, {and} \bibinfo{person}{Yee~Whye Teh}.} \bibinfo{year}{2019}\natexlab{}.
\newblock \showarticletitle{Augmented neural odes}.
\newblock \bibinfo{journal}{\emph{Advances in neural information processing systems}}  \bibinfo{volume}{32} (\bibinfo{year}{2019}).
\newblock


\bibitem[Dutt et~al\mbox{.}(2024)]%
        {dutt2023diffusion}
\bibfield{author}{\bibinfo{person}{Niladri~Shekhar Dutt}, \bibinfo{person}{Sanjeev Muralikrishnan}, {and} \bibinfo{person}{Niloy~J Mitra}.} \bibinfo{year}{2024}\natexlab{}.
\newblock \showarticletitle{Diffusion 3d features (diff3f): Decorating untextured shapes with distilled semantic features}. In \bibinfo{booktitle}{\emph{Proceedings of the IEEE/CVF Conference on Computer Vision and Pattern Recognition}}. \bibinfo{pages}{4494--4504}.
\newblock


\bibitem[Egger et~al\mbox{.}(2019)]%
        {morphableSurvey:19}
\bibfield{author}{\bibinfo{person}{Bernhard Egger}, \bibinfo{person}{William A.~P. Smith}, \bibinfo{person}{Ayush Tewari}, \bibinfo{person}{Stefanie Wuhrer}, \bibinfo{person}{Michael Zollh{\"{o}}fer}, \bibinfo{person}{Thabo Beeler}, \bibinfo{person}{Florian Bernard}, \bibinfo{person}{Timo Bolkart}, \bibinfo{person}{Adam Kortylewski}, \bibinfo{person}{Sami Romdhani}, \bibinfo{person}{Christian Theobalt}, \bibinfo{person}{Volker Blanz}, {and} \bibinfo{person}{Thomas Vetter}.} \bibinfo{year}{2019}\natexlab{}.
\newblock \showarticletitle{3D Morphable Face Models - Past, Present and Future}.
\newblock \bibinfo{journal}{\emph{CoRR}}  \bibinfo{volume}{abs/1909.01815} (\bibinfo{year}{2019}).
\newblock
\showeprint[arXiv]{1909.01815}
\urldef\tempurl%
\url{http://arxiv.org/abs/1909.01815}
\showURL{%
\tempurl}


\bibitem[Fan et~al\mbox{.}(2014)]%
        {fan2014active}
\bibfield{author}{\bibinfo{person}{Ye Fan}, \bibinfo{person}{Joshua Litven}, {and} \bibinfo{person}{Dinesh~K Pai}.} \bibinfo{year}{2014}\natexlab{}.
\newblock \showarticletitle{Active volumetric musculoskeletal systems}.
\newblock \bibinfo{journal}{\emph{ACM Transactions on Graphics (TOG)}} \bibinfo{volume}{33}, \bibinfo{number}{4} (\bibinfo{year}{2014}), \bibinfo{pages}{1--9}.
\newblock


\bibitem[Fragkiadaki et~al\mbox{.}(2015)]%
        {fragkiadaki2015erd}
\bibfield{author}{\bibinfo{person}{K. Fragkiadaki}, \bibinfo{person}{S. Levine}, \bibinfo{person}{P. Felsen}, {and} \bibinfo{person}{J. Malik}.} \bibinfo{year}{2015}\natexlab{}.
\newblock \showarticletitle{Recurrent Network Models for Human Dynamics}. In \bibinfo{booktitle}{\emph{ICCV}}.
\newblock


\bibitem[Fr{\"u}hst{\"u}ck et~al\mbox{.}(2022)]%
        {Fruehstueck2022InsetGAN}
\bibfield{author}{\bibinfo{person}{Anna Fr{\"u}hst{\"u}ck}, \bibinfo{person}{{Krishna Kumar} Singh}, \bibinfo{person}{Eli Shechtman}, \bibinfo{person}{{Niloy J.} Mitra}, \bibinfo{person}{Peter Wonka}, {and} \bibinfo{person}{Jingwan Lu}.} \bibinfo{year}{2022}\natexlab{}.
\newblock \showarticletitle{InsetGAN for Full-Body Image Generation}. In \bibinfo{booktitle}{\emph{CVPR}}. \bibinfo{pages}{7723--7732}.
\newblock


\bibitem[Fu et~al\mbox{.}(2011)]%
        {fzlm_animatedLineDraw_siga_11}
\bibfield{author}{\bibinfo{person}{Hongbo Fu}, \bibinfo{person}{Shizhe Zhjou}, \bibinfo{person}{Ligang Liu}, {and} \bibinfo{person}{Niloy Mitra}.} \bibinfo{year}{2011}\natexlab{}.
\newblock \showarticletitle{Animated Construction of Line Drawings}.
\newblock \bibinfo{journal}{\emph{ACM Transactions on Graphics}} \bibinfo{volume}{30}, \bibinfo{number}{6} (\bibinfo{year}{2011}).
\newblock


\bibitem[Giorgi et~al\mbox{.}(2007)]%
        {giorgi2007shape}
\bibfield{author}{\bibinfo{person}{Daniela Giorgi}, \bibinfo{person}{Silvia Biasotti}, {and} \bibinfo{person}{Laura Paraboschi}.} \bibinfo{year}{2007}\natexlab{}.
\newblock \showarticletitle{Shape retrieval contest 2007: Watertight models track}.
\newblock \bibinfo{journal}{\emph{SHREC competition}} \bibinfo{volume}{8}, \bibinfo{number}{7} (\bibinfo{year}{2007}), \bibinfo{pages}{7}.
\newblock


\bibitem[Harvey et~al\mbox{.}(2020)]%
        {harvey2020robust}
\bibfield{author}{\bibinfo{person}{Félix~G. Harvey}, \bibinfo{person}{Mike Yurick}, \bibinfo{person}{Derek Nowrouzezahrai}, {and} \bibinfo{person}{Christopher Pal}.} \bibinfo{year}{2020}\natexlab{}.
\newblock \showarticletitle{Robust Motion In-Betweening}.
\newblock  \bibinfo{volume}{39}, \bibinfo{number}{4} (\bibinfo{year}{2020}).
\newblock


\bibitem[He et~al\mbox{.}(2022)]%
        {he2022nemf}
\bibfield{author}{\bibinfo{person}{Chengan He}, \bibinfo{person}{Jun Saito}, \bibinfo{person}{James Zachary}, \bibinfo{person}{Holly Rushmeier}, {and} \bibinfo{person}{Yi Zhou}.} \bibinfo{year}{2022}\natexlab{}.
\newblock \showarticletitle{NeMF: Neural Fields for Kinematic Animation}. In \bibinfo{booktitle}{\emph{NeurIPS}}.
\newblock


\bibitem[Holden et~al\mbox{.}(2017)]%
        {phaseFunctioned:17}
\bibfield{author}{\bibinfo{person}{Daniel Holden}, \bibinfo{person}{Taku Komura}, {and} \bibinfo{person}{Jun Saito}.} \bibinfo{year}{2017}\natexlab{}.
\newblock \showarticletitle{Phase-functioned neural networks for character control}.
\newblock \bibinfo{journal}{\emph{ACM Trans. Graph.}} \bibinfo{volume}{36}, \bibinfo{number}{4}, Article \bibinfo{articleno}{42} (\bibinfo{date}{jul} \bibinfo{year}{2017}), \bibinfo{numpages}{13}~pages.
\newblock


\bibitem[Jacobson et~al\mbox{.}(2014)]%
        {jacobson2014skinning}
\bibfield{author}{\bibinfo{person}{Alec Jacobson}, \bibinfo{person}{Zhigang Deng}, \bibinfo{person}{Ladislav Kavan}, {and} \bibinfo{person}{J.~P. Lewis}.} \bibinfo{year}{2014}\natexlab{}.
\newblock \showarticletitle{Skinning: Real-time Shape Deformation}. In \bibinfo{booktitle}{\emph{SIGGRAPH Courses}}.
\newblock


\bibitem[Ju et~al\mbox{.}(2005)]%
        {ju2005meanvalue}
\bibfield{author}{\bibinfo{person}{Tao Ju}, \bibinfo{person}{Scott Schaefer}, {and} \bibinfo{person}{Joe Warren}.} \bibinfo{year}{2005}\natexlab{}.
\newblock \showarticletitle{Mean Value Coordinates for Closed Triangular Meshes}. In \bibinfo{booktitle}{\emph{SIGGRAPH}}.
\newblock


\bibitem[Lewis et~al\mbox{.}(2014)]%
        {lewis2014blendshape}
\bibfield{author}{\bibinfo{person}{J.~P. Lewis}, \bibinfo{person}{Ken Anjyo}, \bibinfo{person}{Taehyun Rhee}, \bibinfo{person}{Mengjie Zhang}, \bibinfo{person}{Fred Pighin}, {and} \bibinfo{person}{Zhigang Deng}.} \bibinfo{year}{2014}\natexlab{}.
\newblock \showarticletitle{Practice and Theory of Blendshape Facial Models}.
\newblock \bibinfo{journal}{\emph{Eurographics State of the Art Reports}} (\bibinfo{year}{2014}).
\newblock


\bibitem[Li et~al\mbox{.}(2021a)]%
        {li2021learning}
\bibfield{author}{\bibinfo{person}{Peizhuo Li}, \bibinfo{person}{Kfir Aberman}, \bibinfo{person}{Rana Hanocka}, \bibinfo{person}{Libin Liu}, \bibinfo{person}{Olga Sorkine-Hornung}, {and} \bibinfo{person}{Baoquan Chen}.} \bibinfo{year}{2021}\natexlab{a}.
\newblock \showarticletitle{Learning Skeletal Articulations with Neural Blend Shapes}.
\newblock \bibinfo{journal}{\emph{ACM Transactions on Graphics (TOG)}} \bibinfo{volume}{40}, \bibinfo{number}{4} (\bibinfo{year}{2021}), \bibinfo{pages}{130}.
\newblock


\bibitem[Li et~al\mbox{.}(2021b)]%
        {def4d}
\bibfield{author}{\bibinfo{person}{Yang Li}, \bibinfo{person}{Hikari Takehara}, \bibinfo{person}{Takafumi Taketomi}, \bibinfo{person}{Bo Zheng}, {and} \bibinfo{person}{Matthias Nießner}.} \bibinfo{year}{2021}\natexlab{b}.
\newblock \showarticletitle{4dcomplete: Non-rigid motion estimation beyond the observable surface.}
\newblock \bibinfo{journal}{\emph{IEEE International Conference on Computer Vision (ICCV)}} (\bibinfo{year}{2021}).
\newblock


\bibitem[Liao et~al\mbox{.}(2022)]%
        {skeletonfree}
\bibfield{author}{\bibinfo{person}{Zhouyingcheng Liao}, \bibinfo{person}{Jimei Yang}, \bibinfo{person}{Jun Saito}, \bibinfo{person}{Gerard Pons-Moll}, {and} \bibinfo{person}{Yang Zhou}.} \bibinfo{year}{2022}\natexlab{}.
\newblock \showarticletitle{Skeleton-free Pose Transfer for Stylized 3D Characters}. In \bibinfo{booktitle}{\emph{European Conference on Computer Vision ({ECCV})}}. {Springer}.
\newblock


\bibitem[Loper et~al\mbox{.}(2015)]%
        {SMPL:2015}
\bibfield{author}{\bibinfo{person}{Matthew Loper}, \bibinfo{person}{Naureen Mahmood}, \bibinfo{person}{Javier Romero}, \bibinfo{person}{Gerard Pons-Moll}, {and} \bibinfo{person}{Michael~J. Black}.} \bibinfo{year}{2015}\natexlab{}.
\newblock \showarticletitle{{SMPL}: A Skinned Multi-Person Linear Model}.
\newblock \bibinfo{journal}{\emph{ACM Trans. Graphics (Proc. SIGGRAPH Asia)}} \bibinfo{volume}{34}, \bibinfo{number}{6} (\bibinfo{date}{Oct.} \bibinfo{year}{2015}), \bibinfo{pages}{248:1--248:16}.
\newblock


\bibitem[Mahmood et~al\mbox{.}(2019)]%
        {AMASS:ICCV:2019}
\bibfield{author}{\bibinfo{person}{Naureen Mahmood}, \bibinfo{person}{Nima Ghorbani}, \bibinfo{person}{Nikolaus~F. Troje}, \bibinfo{person}{Gerard Pons-Moll}, {and} \bibinfo{person}{Michael~J. Black}.} \bibinfo{year}{2019}\natexlab{}.
\newblock \showarticletitle{{AMASS}: Archive of Motion Capture as Surface Shapes}. In \bibinfo{booktitle}{\emph{International Conference on Computer Vision}}. \bibinfo{pages}{5442--5451}.
\newblock


\bibitem[Marsot et~al\mbox{.}(2023)]%
        {marsot2023correspondencefree}
\bibfield{author}{\bibinfo{person}{Mathieu Marsot}, \bibinfo{person}{Rim Rekik}, \bibinfo{person}{Stefanie Wuhrer}, \bibinfo{person}{Jean-Sébastien Franco}, {and} \bibinfo{person}{Anne-Hélène Olivier}.} \bibinfo{year}{2023}\natexlab{}.
\newblock \bibinfo{title}{Correspondence-free online human motion retargeting}.
\newblock
\showeprint[arxiv]{2302.00556}~[cs.CV]


\bibitem[{Mixamo}(2025)]%
        {Mixamo}
\bibfield{author}{\bibinfo{person}{{Mixamo}}.} \bibinfo{year}{2025}\natexlab{}.
\newblock \bibinfo{title}{{Mixamo Animation Service}}.
\newblock
\urldef\tempurl%
\url{https://www.mixamo.com/}
\showURL{%
\tempurl}
\newblock
\shownote{Accessed: 2025-03-08}.


\bibitem[Modi et~al\mbox{.}(2020)]%
        {modi2020efficient}
\bibfield{author}{\bibinfo{person}{V. Modi}, \bibinfo{person}{L. Fulton}, \bibinfo{person}{A. Jacobson}, \bibinfo{person}{S. Sueda}, {and} \bibinfo{person}{D.I.W. Levin}.} \bibinfo{year}{2020}\natexlab{}.
\newblock \showarticletitle{EMU: Efficient Muscle Simulation in Deformation Space}.
\newblock \bibinfo{journal}{\emph{Computer Graphics Forum}} (\bibinfo{date}{Dec} \bibinfo{year}{2020}).
\newblock
\href{https://doi.org/10.1111/cgf.14185}{doi:\nolinkurl{10.1111/cgf.14185}}


\bibitem[Muralikrishnan et~al\mbox{.}(2024)]%
        {trj}
\bibfield{author}{\bibinfo{person}{Sanjeev Muralikrishnan}, \bibinfo{person}{Niladri~Shekhar Dutt}, \bibinfo{person}{Siddhartha Chaudhuri}, \bibinfo{person}{Noam Aigerman}, \bibinfo{person}{Vladimir Kim}, \bibinfo{person}{Matthew Fisher}, {and} \bibinfo{person}{Niloy~J Mitra}.} \bibinfo{year}{2024}\natexlab{}.
\newblock \showarticletitle{Temporal Residual Jacobians For Rig-free Motion Transfer}.
\newblock \bibinfo{journal}{\emph{ECCV 2024 preprint arXiv:2407.14958}} (\bibinfo{year}{2024}).
\newblock


\bibitem[Nicolet et~al\mbox{.}(2021)]%
        {Nicolet2021Large}
\bibfield{author}{\bibinfo{person}{Baptiste Nicolet}, \bibinfo{person}{Alec Jacobson}, {and} \bibinfo{person}{Wenzel Jakob}.} \bibinfo{year}{2021}\natexlab{}.
\newblock \showarticletitle{Large Steps in Inverse Rendering of Geometry}.
\newblock \bibinfo{journal}{\emph{ACM Transactions on Graphics (Proceedings of SIGGRAPH Asia)}} \bibinfo{volume}{40}, \bibinfo{number}{6} (\bibinfo{date}{Dec.} \bibinfo{year}{2021}).
\newblock
\href{https://doi.org/10.1145/3478513.3480501}{doi:\nolinkurl{10.1145/3478513.3480501}}


\bibitem[Park and Hodgins(2008)]%
        {park2008data}
\bibfield{author}{\bibinfo{person}{Sang~Il Park} {and} \bibinfo{person}{Jessica~K Hodgins}.} \bibinfo{year}{2008}\natexlab{}.
\newblock \showarticletitle{Data-driven modeling of skin and muscle deformation}.
\newblock In \bibinfo{booktitle}{\emph{ACM SIGGRAPH 2008 papers}}. \bibinfo{pages}{1--6}.
\newblock


\bibitem[Pons-Moll et~al\mbox{.}(2015)]%
        {pons2015dyna}
\bibfield{author}{\bibinfo{person}{Gerard Pons-Moll}, \bibinfo{person}{Javier Romero}, \bibinfo{person}{Naureen Mahmood}, {and} \bibinfo{person}{Michael~J Black}.} \bibinfo{year}{2015}\natexlab{}.
\newblock \showarticletitle{Dyna: A model of dynamic human shape in motion}.
\newblock \bibinfo{journal}{\emph{ACM Transactions on Graphics (TOG)}} \bibinfo{volume}{34}, \bibinfo{number}{4} (\bibinfo{year}{2015}), \bibinfo{pages}{1--14}.
\newblock


\bibitem[Qi et~al\mbox{.}(2016)]%
        {qi2016pointnet}
\bibfield{author}{\bibinfo{person}{Charles~R Qi}, \bibinfo{person}{Hao Su}, \bibinfo{person}{Kaichun Mo}, {and} \bibinfo{person}{Leonidas~J Guibas}.} \bibinfo{year}{2016}\natexlab{}.
\newblock \showarticletitle{PointNet: Deep Learning on Point Sets for 3D Classification and Segmentation}.
\newblock \bibinfo{journal}{\emph{arXiv preprint arXiv:1612.00593}} (\bibinfo{year}{2016}).
\newblock


\bibitem[Qiao et~al\mbox{.}(2018)]%
        {qiao2018learning}
\bibfield{author}{\bibinfo{person}{Yi-Ling Qiao}, \bibinfo{person}{Lin Gao}, \bibinfo{person}{Yu-Kun Lai}, {and} \bibinfo{person}{Shihong Xia}.} \bibinfo{year}{2018}\natexlab{}.
\newblock \bibinfo{title}{Learning Bidirectional LSTM Networks for Synthesizing 3D Mesh Animation Sequences}.
\newblock
\showeprint[arxiv]{1810.02042}~[cs.GR]


\bibitem[Qin et~al\mbox{.}(2023)]%
        {qin2023NFR}
\bibfield{author}{\bibinfo{person}{Dafei Qin}, \bibinfo{person}{Jun Saito}, \bibinfo{person}{Noam Aigerman}, \bibinfo{person}{Groueix Thibault}, {and} \bibinfo{person}{Taku Komura}.} \bibinfo{year}{2023}\natexlab{}.
\newblock \showarticletitle{Neural Face Rigging for Animating and Retargeting Facial Meshes in the Wild}. In \bibinfo{booktitle}{\emph{SIGGRAPH 2023 Conference Papers}}.
\newblock


\bibitem[Raab et~al\mbox{.}(2023)]%
        {raab2023single}
\bibfield{author}{\bibinfo{person}{Sigal Raab}, \bibinfo{person}{Inbal Leibovitch}, \bibinfo{person}{Guy Tevet}, \bibinfo{person}{Moab Arar}, \bibinfo{person}{Amit~H Bermano}, {and} \bibinfo{person}{Daniel Cohen-Or}.} \bibinfo{year}{2023}\natexlab{}.
\newblock \showarticletitle{Single motion diffusion}.
\newblock \bibinfo{journal}{\emph{arXiv preprint arXiv:2302.05905}} (\bibinfo{year}{2023}).
\newblock


\bibitem[Rombach et~al\mbox{.}(2021)]%
        {rombach2021highresolution}
\bibfield{author}{\bibinfo{person}{Robin Rombach}, \bibinfo{person}{Andreas Blattmann}, \bibinfo{person}{Dominik Lorenz}, \bibinfo{person}{Patrick Esser}, {and} \bibinfo{person}{Björn Ommer}.} \bibinfo{year}{2021}\natexlab{}.
\newblock \bibinfo{title}{High-Resolution Image Synthesis with Latent Diffusion Models}.
\newblock
\showeprint[arxiv]{2112.10752}~[cs.CV]


\bibitem[Rong et~al\mbox{.}(2008)]%
        {rong2008spectral}
\bibfield{author}{\bibinfo{person}{Guodong Rong}, \bibinfo{person}{Yan Cao}, {and} \bibinfo{person}{Xiaohu Guo}.} \bibinfo{year}{2008}\natexlab{}.
\newblock \showarticletitle{Spectral mesh deformation}.
\newblock \bibinfo{journal}{\emph{The Visual Computer}}  \bibinfo{volume}{24} (\bibinfo{year}{2008}).
\newblock


\bibitem[Santesteban et~al\mbox{.}(2020)]%
        {santesteban}
\bibfield{author}{\bibinfo{person}{Igor Santesteban}, \bibinfo{person}{Elena Garces}, \bibinfo{person}{Miguel~A Otaduy}, {and} \bibinfo{person}{Dan Casas}.} \bibinfo{year}{2020}\natexlab{}.
\newblock \showarticletitle{SoftSMPL: Data-driven Modeling of Nonlinear Soft-tissue Dynamics for Parametric Humans}. In \bibinfo{booktitle}{\emph{Computer Graphics Forum}}, Vol.~\bibinfo{volume}{39}. Wiley Online Library, \bibinfo{pages}{65--75}.
\newblock


\bibitem[Sumner and Popovi\'{c}(2004)]%
        {defTransfer:sigg:04}
\bibfield{author}{\bibinfo{person}{Robert~W. Sumner} {and} \bibinfo{person}{Jovan Popovi\'{c}}.} \bibinfo{year}{2004}\natexlab{}.
\newblock \showarticletitle{Deformation Transfer for Triangle Meshes}.
\newblock \bibinfo{journal}{\emph{ACM Trans. Graph.}} \bibinfo{volume}{23}, \bibinfo{number}{3} (\bibinfo{year}{2004}), \bibinfo{pages}{399–405}.
\newblock


\bibitem[Sun et~al\mbox{.}(2019)]%
        {hrnet}
\bibfield{author}{\bibinfo{person}{Ke Sun}, \bibinfo{person}{Bin Xiao}, \bibinfo{person}{Dong Liu}, {and} \bibinfo{person}{Jingdong Wang}.} \bibinfo{year}{2019}\natexlab{}.
\newblock \showarticletitle{Deep High-Resolution Representation Learning for Human Pose Estimation}. In \bibinfo{booktitle}{\emph{CVPR}}.
\newblock


\bibitem[Sun and Murata(2020)]%
        {cafm}
\bibfield{author}{\bibinfo{person}{Yifan Sun} {and} \bibinfo{person}{Noboru Murata}.} \bibinfo{year}{2020}\natexlab{}.
\newblock \showarticletitle{CAFM: A 3D Morphable Model for Animals}. In \bibinfo{booktitle}{\emph{2020 IEEE Winter Applications of Computer Vision Workshops (WACVW)}}. \bibinfo{pages}{20--24}.
\newblock
\href{https://doi.org/10.1109/WACVW50321.2020.9096941}{doi:\nolinkurl{10.1109/WACVW50321.2020.9096941}}


\bibitem[Xu et~al\mbox{.}(2020)]%
        {RigNet:20}
\bibfield{author}{\bibinfo{person}{Zhan Xu}, \bibinfo{person}{Yang Zhou}, \bibinfo{person}{Evangelos Kalogerakis}, \bibinfo{person}{Chris Landreth}, {and} \bibinfo{person}{Karan Singh}.} \bibinfo{year}{2020}\natexlab{}.
\newblock \showarticletitle{RigNet: Neural Rigging for Articulated Characters}.
\newblock \bibinfo{journal}{\emph{ACM Trans. on Graphics}}  \bibinfo{volume}{39} (\bibinfo{year}{2020}).
\newblock


\bibitem[Yifan et~al\mbox{.}(2020)]%
        {Yifan:NeuralCage:2020}
\bibfield{author}{\bibinfo{person}{Wang Yifan}, \bibinfo{person}{Noam Aigerman}, \bibinfo{person}{Vladimir~G. Kim}, \bibinfo{person}{Siddhartha Chaudhuri}, {and} \bibinfo{person}{Olga Sorkine-Hornung}.} \bibinfo{year}{2020}\natexlab{}.
\newblock \showarticletitle{Neural Cages for Detail-Preserving 3D Deformations}. In \bibinfo{booktitle}{\emph{CVPR}}.
\newblock


\bibitem[Zhang et~al\mbox{.}(2023)]%
        {zhang2023skinned}
\bibfield{author}{\bibinfo{person}{Jiaxu Zhang}, \bibinfo{person}{Junwu Weng}, \bibinfo{person}{Di Kang}, \bibinfo{person}{Fang Zhao}, \bibinfo{person}{Shaoli Huang}, \bibinfo{person}{Xuefei Zhe}, \bibinfo{person}{Linchao Bao}, \bibinfo{person}{Ying Shan}, \bibinfo{person}{Jue Wang}, {and} \bibinfo{person}{Zhigang Tu}.} \bibinfo{year}{2023}\natexlab{}.
\newblock \showarticletitle{Skinned motion retargeting with residual perception of motion semantics \& geometry}. In \bibinfo{booktitle}{\emph{Proceedings of the IEEE/CVF Conference on Computer Vision and Pattern Recognition}}. \bibinfo{pages}{13864--13872}.
\newblock


\bibitem[Zhou et~al\mbox{.}(2020)]%
        {zhou20unsupervised}
\bibfield{author}{\bibinfo{person}{Keyang Zhou}, \bibinfo{person}{Bharat~Lal Bhatnagar}, {and} \bibinfo{person}{Gerard Pons-Moll}.} \bibinfo{year}{2020}\natexlab{}.
\newblock \showarticletitle{Unsupervised Shape and Pose Disentanglement for 3D Meshes}. In \bibinfo{booktitle}{\emph{European Conference on Computer Vision (ECCV)}}.
\newblock


\end{thebibliography}

\end{document}